\documentclass[a4paper,fleqn,usenatbib]{mnras}
\usepackage{footmisc}
\usepackage{graphicx}
\usepackage{amssymb}
\usepackage{amsmath}
\usepackage{wasysym}
\usepackage{color}
\usepackage[normalem]{ulem}
\usepackage{psfrag}
\usepackage{multirow,bigdelim}
\graphicspath{{./}{../plots/}}

\newcommand{\halpha}{\hbox{H$\alpha$}}
\newcommand{\hbeta}{\hbox{H$\beta$}}

\newcommand{\heii}{\hbox{\ion{He}{ii}~$\lambda 4686$}}
\newcommand{\oiii}{\hbox{[\ion{O}{iii}]~$\lambda 5007$}}

\newcommand{\asli}{\hbox{ASASSN-14li}}

\newcommand{\ea}{\hbox{E$+$A}}

\newcommand{\mbh}{\hbox{$M_{\rm BH}$}}
\newcommand{\msun}{\hbox{M$_{\odot}$}}

\newcommand{\Ngal}{\hbox{$N^{\rm Gal}_{H}$}}

\newcommand{\SWIFT}{\textit{Swift}}
 
\newcommand{\IRAF}{\textsc{iraf}}

\newcounter{minirefcount}
\title[The Evolution of \asli]{The Long Term Evolution of \asli}

\author[J. S. Brown et al.]{J. S. Brown,$^{1}$\thanks{E-mail: brown@astronomy.ohio-state.edu}
T. W.-S Holoien,$^{1,2}$
K. Auchettl,$^{2,3}$
K. Z. Stanek,$^{1,2}$
C. S. Kochanek,$^{1,2}$ \newauthor
B. J. Shappee,$^{4,5,6}$
J. L. Prieto,$^{7,8}$
and D. Grupe$^{9}$  
\\
$^{1}$ Department of Astronomy, The Ohio State University, 140 West 18th Avenue, Columbus, OH 43210, USA\\
$^{2}$ Center for Cosmology and Astro-Particle Physics, The Ohio State University, 191 West Woodruff Avenue, Columbus, OH 43210, USA\\
$^{3}$ Department of Physics, The Ohio State University, 191 W. Woodruff Avenue, Columbus, OH 43210, USA\\
$^{4}$ Carnegie Observatories, 813 Santa Barbara Street, Pasadena, CA 91101, USA\\
$^{5}$ Hubble Fellow\\
$^{6}$ Carnegie-Princeton Fellow\\
$^{7}$ N\'ucleo de Astronom\'ia de la Facultad de Ingenier\'ia, Universidad Diego Portales, Av. Ej\'ercito 441, Santiago, Chile \\
$^{8}$ Millennium Institute of Astrophysics, Santiago, Chile \\
$^{9}$ Department of Earth and Space Science, Morehead State University, 235 Martindale Dr., Morehead, KY 40351, USA
}

\date{Accepted XXX. Received YYY; in original form ZZZ}

\pubyear{2016}

\begin{document}
\label{firstpage}
\pagerange{\pageref{firstpage}--\pageref{lastpage}}
\maketitle

\begin{abstract}
We present late-time optical spectroscopy taken with the Large Binocular Telescope's Multi-Object Double Spectrograph, late-time \SWIFT\ UVOT and XRT observations, as well as improved ASAS-SN pre-discovery limits on the nearby ($d=90.3$ Mpc, $z=0.0206$) tidal disruption event (TDE) \asli. The late-time optical spectra show \halpha\ emission well in excess of that seen in the SDSS host galaxy spectrum, indicating that the processes powering the luminous flares associated with TDEs can operate for several hundreds of days. The \SWIFT\ observations reveal the presence of lingering apparently thermal UV (T$_{\rm UV} \sim 3.5\times10^4$~K) and X-ray (T$_{\rm X} \sim 7\times10^5$~K) emission. The characteristic temperatures evolve by, at most, a factor of $\sim2$ over the 600 day follow-up campaign. The X-ray, UV, and \halpha\ luminosities evolve roughly in tandem and at a rate that is consistent with a power-law decay at late times. This behavior is in stark contrast with the majority of optically discovered TDEs, which are X-ray faint and evolve on shorter timescales. Finally we address how the unique properties of \asli\ can be used to probe the relationship between the TDE rate and host galaxy properties.
\end{abstract}


\section{Introduction}
\label{sec:intro}

Near the centers of galaxies, stars can make close approaches to supermassive black holes (SMBHs). Roughly speaking, if the pericenter of a star's orbit is outside the SMBH horizon but inside the Roche limit, the star will be disrupted. When a main sequence star is disrupted, approximately half of the stellar debris will remain on bound orbits and asymptotically return to pericenter at a rate proportional to $t^{-5/3}$ \citep{Rees88,Evans89,Phinney89}. The observational consequences of these tidal disruption events (TDEs) are varied and depend on the physical properties of the disrupted star \citep[e.g.][]{MacLeod12,Kochanek16_stellar}, the post-disruption evolution of the accretion stream \citep[e.g.][]{Kochanek94,Strubbe09,Guillochon13,Hayasaki13,Hayasaki16,Piran15,Shiokawa15}, and complex radiative transfer effects \citep[e.g.][]{Gaskell14,Strubbe15,Roth16}. 

A large sample of well studied TDEs is crucial for understanding the observational characteristics and physical processes governing these exotic objects. While the number of well studied TDE candidates is growing \citep[e.g.][]{vanVelzen11, Cenko12, Gezari12, Arcavi14, Chornock14, Holoien14, Gezari15, Vinko15, Holoien16_15oi, Holoien16_14li, Brown16b}, the diversity of the observational signatures is surprising given that observed TDEs should be heavily dominated by stars and SMHBs spanning a narrow parameter range \citep{Kochanek16_demo}. Perhaps most notably, the majority of optically discovered TDEs show little evidence of X-ray emission, while the energetics of other TDE candidates may be dominated by their X-ray emission  \citep[e.g.][]{Grupe99,KomossaBade99,KomossaGreiner99}.

ASASSN-14li \citep{Jose14,Holoien16_14li} was a nearby ($d\sim$~90 Mpc, $z=0.0206$) TDE discovered by the All-Sky Automated Survey for SuperNovae \citep[ASAS-SN;][]{Shappee14} on 2014-11-22.6 (MJD~=~56983.6). An immediate follow-up campaign \citep{Holoien16_14li} observed \asli\ for $\sim$~200 days. We gathered data from a wide variety of both ground and space based observatories and found that the spectral characteristics of \asli\ resembled the ``intermediate H$+$He'' TDEs from \citet{Arcavi14}, while the optical/NUV evolution was consistent with that of a blackbody and a roughly exponentially declining luminosity. We also found significant spectral evolution: at early times the \heii\ feature is dominant, while at later times it is merely comparable in strength to the Balmer lines. This is in contrast with ASASSN-14ae, in which the \heii\ line became stronger relative to the Balmer lines as the event progressed \citep{Brown16b}. Unlike the two other ASASSN TDEs, \asli\ showed strong X-ray emission, and, due to it's proximity, was the target of several ground-based \citep{Alexander16,vanVelzen16,RomeroCanizales16}, space-based \citep{Miller15,Cenko16,Peng16,Jiang16}, and theoretical efforts \citep{Krolik16,Kochanek16_stellar}.

In this paper we follow the evolution of \asli\ to $\sim600$ days after discovery. We present improved ASAS-SN pre-discovery upper limits, late-time optical spectra taken with the Multi-Object Double Spectrograph 1 (MODS1) on the 8.4 m Large Binocular Telescope (LBT), and extensive UVOT and XRT observations from the \SWIFT\ space telescope, which provide unprecedented insight into this rare class of objects. In Section~\ref{sec:data} we describe our observations, in Section~\ref{sec:discussion} we present our measurements of the late-time evolution, and finally in Section~\ref{sec:conclusions} we provide a summary of our results and discuss the implications for future studies.

\section{Observations}
\label{sec:data}

In this section we summarize the optical, UV, and X-ray observations taken during our $\sim600$ day follow-up campaign of \asli.

\subsection{Spectroscopic Observations}
\begin{table*}
\caption{LBT/MODS1 observations. \label{tab:tab1}}

\begin{minipage}{\textwidth}
\begin{tabular}{@{}lrrrcrcrrrrrrrrrrrrr}
\hline
\hline

{} & {} & \multicolumn{1}{c}{HJD} & \multicolumn{1}{c}{Pos. Angle} & \multicolumn{1}{c}{Parallactic Angle} & {} & \multicolumn{1}{c}{Flux} & \multicolumn{1}{c}{Seeing} & \multicolumn{1}{c}{Exposure} & \multicolumn{1}{c}{$r'$}\\
\multicolumn{1}{c}{UT Date} & \multicolumn{1}{c}{Day} & \multicolumn{1}{c}{$-2,400,000$} & \multicolumn{1}{c}{[deg]} & \multicolumn{1}{c}{[deg]} & \multicolumn{1}{c}{Airmass} & \multicolumn{1}{c}{Standard} & \multicolumn{1}{c}{[arcsec]} & \multicolumn{1}{c}{[s]$\times$N} & \multicolumn{1}{c}{[mag]}\\

\hline

2015 Jan 20.53 & 58.9  & 57042.53 \hphantom{.}& $-62.0$ \hphantom{.}& $12.0$ to $32.0$ & 1.04 -- 1.05 & Feige 34  & 1.2 \hphantom{..}&  600.0$\times$3 & $15.48 \pm 0.01$\\
2015 Feb 16.36 & 85.7  & 57069.36 \hphantom{.}& $-55.0$ \hphantom{.}& $-39.3$ to $-52.7$ & 1.06 -- 1.13 & Feige 34  & 1.4 \hphantom{..}&  900.0$\times$4 & $15.48 \pm 0.01$\\
2015 May 19.30 & 177.7 & 57161.30 \hphantom{.}& $-125.0$ \hphantom{.}& $-119.2$ to $-122.8$ & 1.22 -- 1.38 & BD+33d2642  & 1.3 \hphantom{..}&  1200.0$\times$3 & $15.53 \pm 0.01$\\
2015 Dec 09.50 & 381.9 & 57365.50 \hphantom{.}& $119.0$ \hphantom{.}& $118.2$ to $120.1$ & 1.30 -- 1.51 & Feige 67  & 0.8 \hphantom{..}&  1200.0$\times$3 & $15.53 \pm 0.01$\\
2016 Feb 08.37 & 442.8 & 57426.36 \hphantom{.}& $105.0$ \hphantom{.}& $122.8$ to $130.7$ & 1.11 -- 1.20 & G191-B2B  & 1.0 \hphantom{..}&  1200.0$\times$3 & $15.53 \pm 0.01$\\
2016 Apr 04.17 & 498.6 & 57482.17 \hphantom{.}& $125.0$ \hphantom{.}& $119.2$ to $135.1$ & 1.09 -- 1.41 & Feige 67  & 1.0 \hphantom{..}&  1200.0$\times$6 & $15.53 \pm 0.01$\\

\hline
\end{tabular}

\medskip

The Day column is in days since discovery ($t_{disc} = 56983.6$).\\

\end{minipage}
\end{table*}

Follow-up spectroscopy of \asli\ was obtained with MODS1 \citep{Pogge10} on the LBT between January 2015 and April 2016. The observations were performed in longslit mode with a 1\farcs2 slit. MODS1 uses a dichroic that splits the light into separately optimized red and blue channels at $\sim$~5650\,\AA. The blue CCD covers a  wavelength range of $\sim$~3200 -- 5650\,\AA, with a spectral resolution of 2.4\,\AA, while the red CCD covers a wavelength range of $\sim$~5650 -- 10000\,\AA, with a spectral resolution of 3.4\,\AA.

Our first spectrum was taken on 2015-01-20 ($t = 58.9$ days after discovery; \citealt{Jose14}), and consisted of three 600s exposures. The observations on 2015-02-16 ($t = 85.8$ days) consisted of four 900s exposures. The following three observations ($t = 177.7, 381.9$, and $442.8$ days) consisted of three 1200s exposures. Our most recent, and deepest, observation ($t = 498.6$ days) consisted of six 1200s exposures. The position angle of the slit was chosen to match the parallactic angle at the midpoint of the observations in order to minimize slit losses due to differential atmospheric refraction \citep{Filippenko82}. 

We obtained bias frames and Hg(Ar), Ne, Xe, and Kr calibration lamp images for wavelength calibration. If the arc lamp or flat field data were not available on the night of the observation, we used calibration data obtained within 1-2 days of our observations. Given the stability of MODS1 over the course of an observing run, this is sufficient to provide accurate calibrations. Night sky lines were used to correct for the small ($\sim$~1\,\AA) residual flexure. Standard stars were observed with a 5$\times$60\arcsec\ spectrophotometric slit mask and used to calibrate the response curve. The standard stars are from the HST Primary Calibrator list, which is composed of well observed northern-hemisphere standards from the lists of \citet{Oke90} and \citet{Bohlin95}. We list the information regarding our observational dates and configurations in Table~\ref{tab:tab1}.

The \textit{modsCCDRed}\footnote{\url{http://www.astronomy.ohio-state.edu/MODS/Software/modsCCDRed/}} \textsc{python} package was used to bias subtract, flat field, and illumination correct the raw data frames. We removed cosmic rays with \textit{L.A.Cosmic} \citep{vanDokkum01}. The sky subtraction and one-dimensional extraction were performed with the \textit{modsIDL} pipeline\footnote{\url{http://www.astronomy.ohio-state.edu/MODS/Software/modsIDL/}}. We correct residual sky features with reduced spectra of standard stars observed on the same night under similar conditions using the \IRAF\ task \textit{telluric}. Finally, we combined the individual exposures from each epoch, yielding a total of six high S/N spectra corresponding to the six observation epochs.

In order to facilitate comparison of spectra across multiple observing epochs, we flux-calibrated each spectra with the contemporaneous $r'$-band MODS acquisition images. Following \citet{Shappee13} and \citet{Brown16b}, We performed aperture photometry on the \asli\ host and bright stars in the field with the \IRAF\ package \textit{apphot}. We determined the $r'$ magnitude scale factor for stars in the field and used this to both calibrate the broadband flux of \asli\ in the acquisition image and the new spectra. The typical uncertainty in the aperture photometry is $\sim0.01$ magnitudes. The spectral evolution of \asli\ is presented in Section~\ref{sec:spec}.

\subsection{\SWIFT\ Observations}

After the publication of \citet{Holoien16_14li}, we also obtained additional \SWIFT\ observations of \asli. The UVOT \citep{Poole08} observations were obtained in six filters: $V$ (5468 \AA), $B$ (4392 \AA), $U$ (3465 \AA), $UVW1$ (2600 \AA), $UVM2$ (2246 \AA), and $UVW2$ (1928 \AA). We used the UVOT software task \textsc{uvotsource} to extract the source counts from a 5\farcs0 radius region and a sky region with a radius of $\sim$~40\arcsec. The UVOT count rates were converted into magnitudes and fluxes based on the most recent UVOT calibration \citep{Poole08, Breeveld10}. The uncorrected UVOT magnitudes are presented in Table~\ref{tab:phot}.

We simultaneously obtained Swift X-ray Telescope \citep[XRT;][]{Burrows05} observations of the source. The XRT was run in photon-counting (PC) mode \citep{Hill04} which is the standard imaging mode of the XRT. We reduced all observations following the \SWIFT\ XRT data reduction guide\footnote{\url{http://swift.gsfc.nasa.gov/analysis/xrt\_swguide\_v1\_2.pdf}} and reprocessed the level one XRT data using the \SWIFT\ \textit{xrtpipeline} version 0.13.2 script, producing cleaned event files and exposure maps for each observation.

To extract the number of background subtracted source counts in the 0.3--10.0~keV energy band from each individual observation, we used a source region centered on the position of ASASSN-14li with a radius of 47$\arcmin$ and a source free background region centered at $(\alpha,\delta)=(12^{h}48^{m}39^{s}, +17^{\circ}46'54'')$ with a radius of 236$\arcmin$. The count rates are presented in Table~\ref{tab:phot} and have not been corrected for Galactic absorption.

To increase the S/N of our observations, we also combined individual observations using \textit{XSELECT} version 2.4c. We first combined the observations into 8 time bins spanning our campaign. Additionally, in order to determine the evolution in the X-ray emission between early and late times, we combined the individual observations into early (first 200 days) and late (remaining $\sim360$ days) time bins, yielding high S/N spectra we can compare directly.

For the merged observations we used the task \textit{xrtproducts} version 0.4.2 to extract both source and background spectra using the same regions used to extract our count rates. The task \textit{xrtproducts} task takes advantage of \textit{XSELECT}'s ability to extract spectra, while using the task \textit{xrtmkarf} and the exposure maps produced during by the \textit{xrtpipeline} to create ancillary response files (ARF) for each spectra. To produce the ARF for the merged event files, we first had to merge their individual exposure maps using \textit{XIMAGE} version 4.5.1 before running \textit{xrtmkarf}. The response matrix files (RMFs) are ready-made files which we were obtained from the most recent calibration database. None of our observations suffer from significant pileup issues. 

To analyse the spectral data, we used the X-ray spectral fitting package (XSPEC) version 12.9.0d and chi-squared statistics. Each individual spectrum was grouped with a minimum of 10 counts per energy bin using the FTOOLS command \textit{grppha}, while the spectra obtained from the merged observations were grouped with a minimum of 20 counts per energy bin. The X-ray observations favor a thermal spectrum over the power-law model used in \citet{Holoien16_14li}. We fit the spectra from 0.3--1.0~keV using an absorbed blackbody spectrum emitted at the redshift of the TDE and integrate the flux in the 0.3--10.0~keV band. We assumed the \citet{Wilms00} abundance model and initially let $N_{H}$, the temperature of the blackbody (kT) and its normalization be free. For a number of the individual observations, especially the later observations, the S/N is insufficient to constrain $N_{H}$, and thus we fix it to 1.64$\times10^{20}$ cm$^{-2}$ which is the Galactic H\textsc{I} column density (\Ngal) in the direction of ASASSN-14li \citep{Kalberla05}.

\subsection{ASAS-SN Pre-Discovery Upper Limits}

To further constrain the early-time light curve of \asli\ we re-evaluated the pre-discovery ASAS-SN non-detection discussed in \citet{Holoien16_14li}. The last ASAS-SN epoch before discovery was observed on 2014-07-13.25 under moderate conditions by the quadruple 14-cm ``Brutus'' telescope in Haleakala, Hawaii. This ASAS-SN field was processed by the standard ASAS-SN pipeline (Shappee et al. in prep.) using the \textit{isis} image subtraction package \citep{Alard98,Alard00}, except we did not allow images with flux from \asli\ to be used in the construction of the reference image. We then performed aperture photometry at the location of \asli\ on the subtracted images using the \textsc{iraf} \textit{apphot} package and calibrated the results using the AAVSO Photometric All-Sky Survey \citep[APASS;][]{Henden16}. There was no excess flux detected at the location of \asli\ over the reference image on 2014-07-13.25 ($t=-132.35$ days), and we place a 3-sigma limit of $V > 17.37$ mag on \asli\ at this epoch. Additionally, we stack the previous 4 epochs taken under favorable conditions (2014-06-18.30 through 2014-06-27.28 or $t=-148.32$ through $-157.30$ days) to place a deeper limit $V > 19.00$ mag during this time. 

\section{Evolution of the Late-Time Emission}
\label{sec:discussion}

In this section we discuss the optical, UV, and X-ray evolution of \asli. We interpret our observations in the context of physically motivated models to gain insight into how the physical conditions evolve over time.

\subsection{Evolution of the Optical Spectra}
\label{sec:spec}

\begin{table*} 
\begin{minipage}{\textwidth} 
\centering
\caption{Measurements of \halpha\ Properties \label{tab:tab2}} 

\begin{tabular}{@{}lcccccc} 
\hline 
\hline 
{} & {} & {$\Delta v$} & {FWHM} & {\halpha\ Flux} & {\halpha\ Luminosity} \\ 
{UT Date} & {Day} & {[$10^3$ km s$^{-1}$]} & {[$10^3$ km s$^{-1}$]} & {[$10^{-14}$ ergs s$^{-1}$ cm$^{-2}$]} & {[$10^{40}$ ergs s$^{-1}$]} \\ 
\hline 

{}2015 Jan 20.53 & \hphantom{ }58.9  & $ -0.10 $ & $  1.82 $ & $ 3.533 \pm 0.008 $ & $3.447 \pm 0.008 $ \\ 
{}2015 Feb 16.40 & \hphantom{ }85.8 & $ -0.11 $ & $  1.68 $ & $ 2.499 \pm 0.007 $ & $2.438 \pm 0.007 $ \\ 
{}2015 May 19.31 & 177.7 & $ -0.25 $ & $  1.42 $ & $ 0.901 \pm 0.004 $ & $0.879 \pm 0.004 $ \\ 
{}2015 Dec 09.51 & 381.9 & $ -0.25 $ & $  1.32 $ & $ 0.530 \pm 0.005 $ & $0.518 \pm 0.005 $ \\ 
{}2016 Feb 08.40 & 442.8 & $ -0.22 $ & $  1.27 $ & $ 0.420 \pm 0.005 $ & $0.410 \pm 0.005 $ \\ 
{}2016 Apr 04.25 & 498.6 & $ -0.24 $ & $  1.08 $ & $ 0.371 \pm 0.005 $ & $0.362 \pm 0.005 $ \\ 
\hline 
\end{tabular} 

\medskip 

\end{minipage} 
\end{table*}

\begin{figure*}
\centering{\includegraphics[scale=1.,width=\textwidth,trim=0.pt 0.pt 0.pt 0.pt,clip]{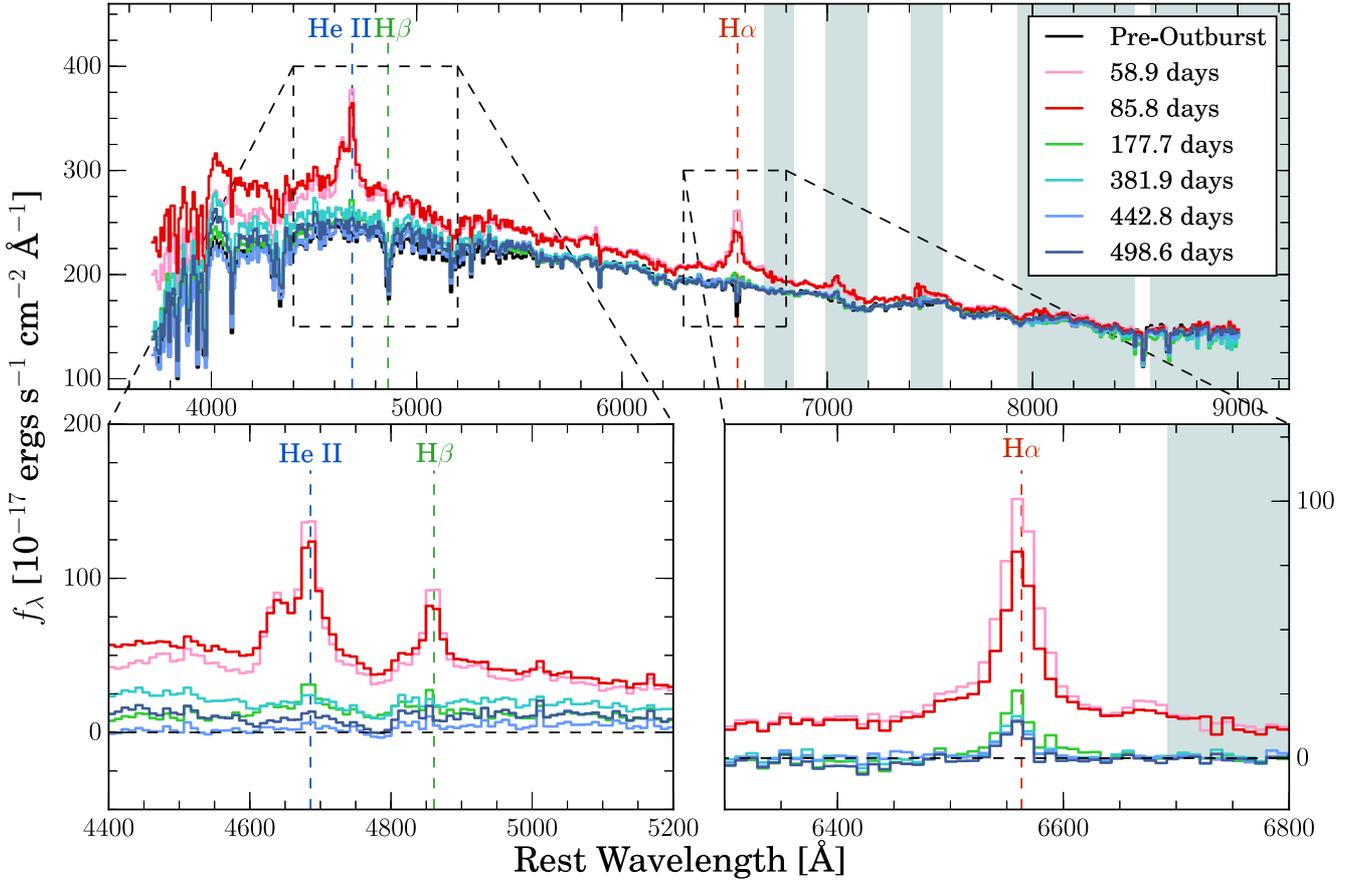}}
\caption{Rest frame flux-calibrated spectra of \asli. Color denotes days since discovery. The top panel shows the full optical spectrum, while the bottom panels show host-subtracted regions in the immediate vicinity of \heii\ (left) and \halpha\ (right). The shaded regions show the location of telluric features where systematic errors may be significant. The spectra show clear temporal evolution in the sense of decreasing continuum and emission line features with increasing time. Even the latest spectra show excess broad emission relative to the host spectrum.}
\label{fig:specEvol}
\end{figure*}

In \citet{Holoien16_14li} we obtained spectra spanning 145 days between UT 2014 December 02 and 2015 April 14. The optical spectra of \asli\ qualitatively resembled that of ASASSN-14ae and the ``intermediate H+He events'' from \citet{Arcavi14}. The spectroscopic evolution mirrored that of ASASS-14ae, in the sense that both showed a weakening blue continuum accompanied by weaker and narrower emission lines with time. However, the emission features of \asli\ were not identical to those of ASASSN-14ae. Specifically, \asli\ showed strong \heii\ emission even at early epochs. The emission line profiles in \asli\ were also narrower and evolved more slowly than in ASASSN-14ae. While part of these differences may be attributed to differences in the ages of the TDEs, there are likely to be additional factors responsible for the observed differences between the two objects.

Figure~\ref{fig:specEvol} shows the evolution of the optical spectra beginning $\sim$~60 days after discovery (pink) and ending $\sim$~500 days after discovery (dark blue). The black spectrum shows the archival SDSS DR7 \citep{Abazajian09} spectrum of the host galaxy taken on 2008-02-15. The top panel shows the full optical spectrum, while the bottom panels show expanded, host-subtracted views of the \heii\ (left) and \halpha\ (right) regions. Prominent spectral features are labeled, while the shaded bands denote regions prone to systematic errors related to telluric correction.

We focus on the evolution of the \halpha\ profile, since it shows the clearest detection at late times and is less prone to systematic errors associated with flux calibration (since we scale the flux to match the $r'$-band photometry). In order to measure the \halpha\ emission from the TDE, we must first subtract the underlying host galaxy. Previous late-time studies of TDEs simply obtained a late-time spectrum of the host after all signatures of the TDE had faded \citep[e.g.][]{Gezari15,Brown16b}. However, in the case of \asli, we still find evidence for residual emission from the TDE even after 500 days, so we must instead subtract the archival SDSS spectrum.

Our initial spectrum taken 60 days after discovery (pink) shows moderately broad (FWHM $\sim$~2000 km s$^{-1}$) \heii, \hbeta, and \halpha\ emission lines superimposed on a blue continuum. The lines are slightly redshifted with respect to the host galaxy ($\Delta v\sim$~100 km s$^{-1}$). The relative weakness of the blue continuum in this spectrum appears to be due to slit losses caused by a combination of variable seeing and slit positioning. If this variability in the continuum were real then it would be evident in the UVOT photometry (see Section~\ref{sec:uvot}), but we find no evidence for such a correlation. The next spectrum, taken $\sim 85$ days after discovery shows slightly weaker, but still prominent \heii, \hbeta, and \halpha\ emission features. Our subsequent spectra taken at $\sim$ 180, 380, 440, and 500 days show a progression in the sense of weakening emission line strength with increased time after discovery. In general, the blue continuum associated with the TDE is weaker at later time. It is not clear if the late-time variability in the blue continuum (Figure~\ref{fig:specEvol}, bottom left panel) is real, but it would be consistent with the apparent $\sim 0.1$ mag variability seen in the UVOT photometry at later times (see Section~\ref{sec:uvot}).

While the \heii\ and \hbeta\ detections are fairly marginal at later times, there is still an unambiguous detection of \halpha\ 500 days after discovery. This suggests that while some TDEs are short lived \citep{Vinko15,Holoien16_15oi, Brown16b}, others may persist long after the time of disruption, even at optical wavelengths. The residual emission lines are unlikely to be attributable to systematics associated with differences in the spatial sampling of the two instruments. The primary evidence for this is that the line profile is broad ($\sim 1000$ km s$^{-1}$) and shows a redward offset relative to the location of \halpha\ in the rest frame of the host galaxy, similar to previous epochs. As an additional check, we performed multiple slit extractions and found no evidence that the \halpha\ emission is a host-subtraction artifact. Finally, the presence of residual \halpha\ emission is fully consistent with our late-time UVOT photometry (see Section~\ref{sec:uvot}). 

We do however measure a slight excess of \oiii\ in our host subtracted spectra, suggesting that the late time MODS spectra include nebular emission not associated with the TDE. Similarly, we find evidence for excess \halpha\ emission corresponding to the narrow nebular \halpha\ feature in the archival spectrum. This is not particularly surprising given the complex morphology of the nebular emission in this galaxy \citep{Prieto16}, and the differences in the spatial sampling of the MODS1 longslit and the SDSS fiber. We estimate the magnitude of this contamination by modeling the excess \halpha\ emission with two components: a broad component corresponding to the emission associated with the TDE, and a narrow ($\sim100$ km s$^{-1}$) component scaled from the \oiii\ residual feature. We find that even in the latest epoch when the \halpha\ flux is weakest, the flux in the narrow component is $\lesssim 20\%$ that of the broad component, and thus does not substantially affect our results.

\subsection{Evolution of the UVOT Photometry}
\label{sec:uvot}

\begin{figure*}
\centering{\includegraphics[scale=1.,width=\textwidth,trim=0.pt 0.pt 0.pt 0.pt,clip]{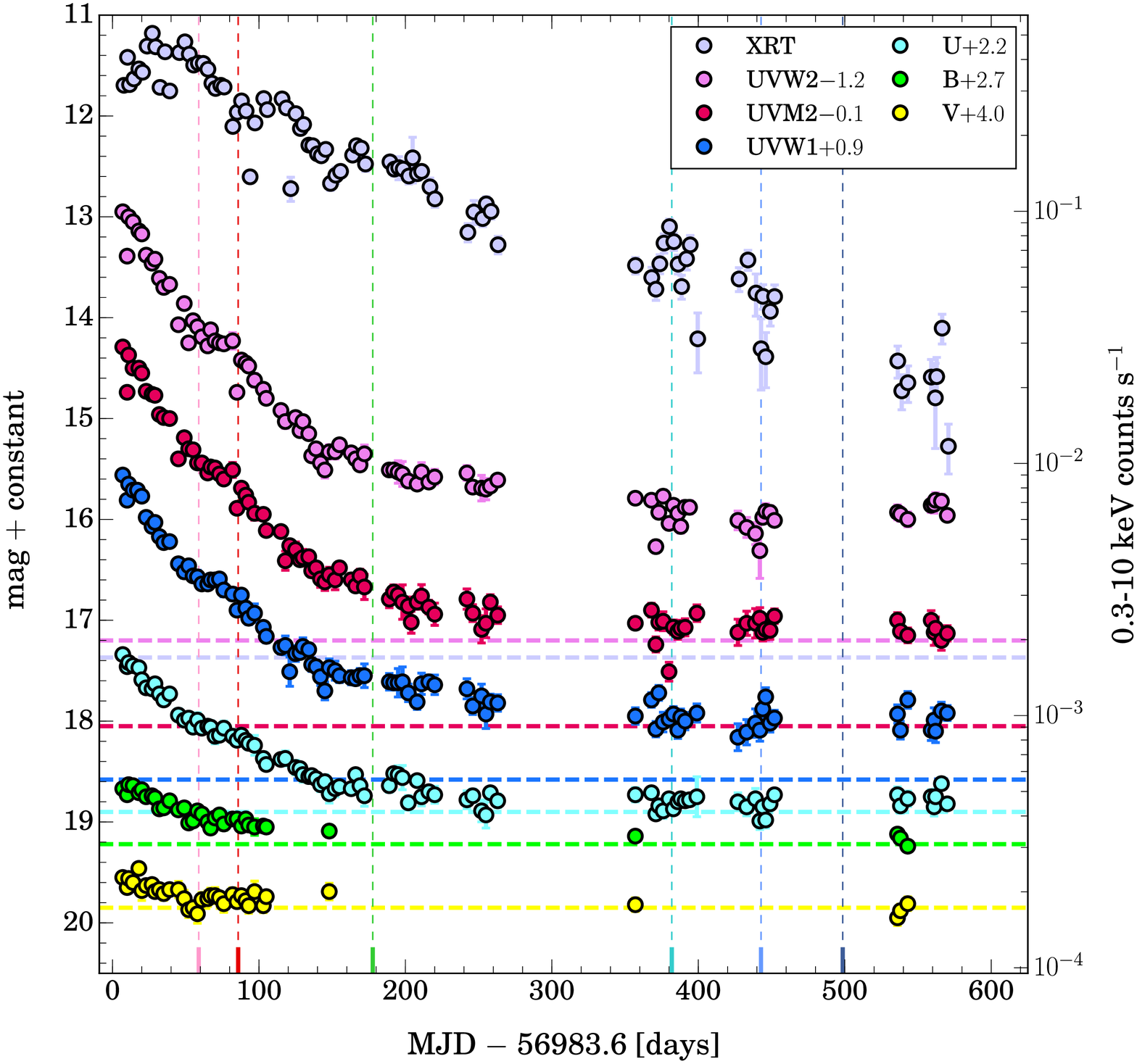}}
\caption{Evolution of \asli\ in the SWIFT UVOT bands from discovery to $\sim 600$ days after discovery. Circles show the observed non-host-subtracted magnitudes. All UV and optical magnitudes are shown in the Vega system (left scale), and X-ray count rates are shown as counts s$^{-1}$ in the 0.3--10.0~keV energy range (right scale). Both scales span the same dynamic range. Horizontal dashed lines show the estimated host magnitudes synthesized from the best fit SED relative to the UVOT magnitudes and the archival X-ray upper limit relative to the X-ray counts. Vertical marks along the time-axis show the dates of our spectroscopic observations. Even after 600 days, \asli\ remains significantly brighter than the host in the UV and X-ray, but is essentially undetectable in the optical bands after $\sim 200$ days. }
\label{fig:swiftLC}
\end{figure*}

\begin{figure}
\centering{\includegraphics[scale=1.,width=0.5\textwidth,trim=0.pt 0.pt 0.pt 0.pt,clip]{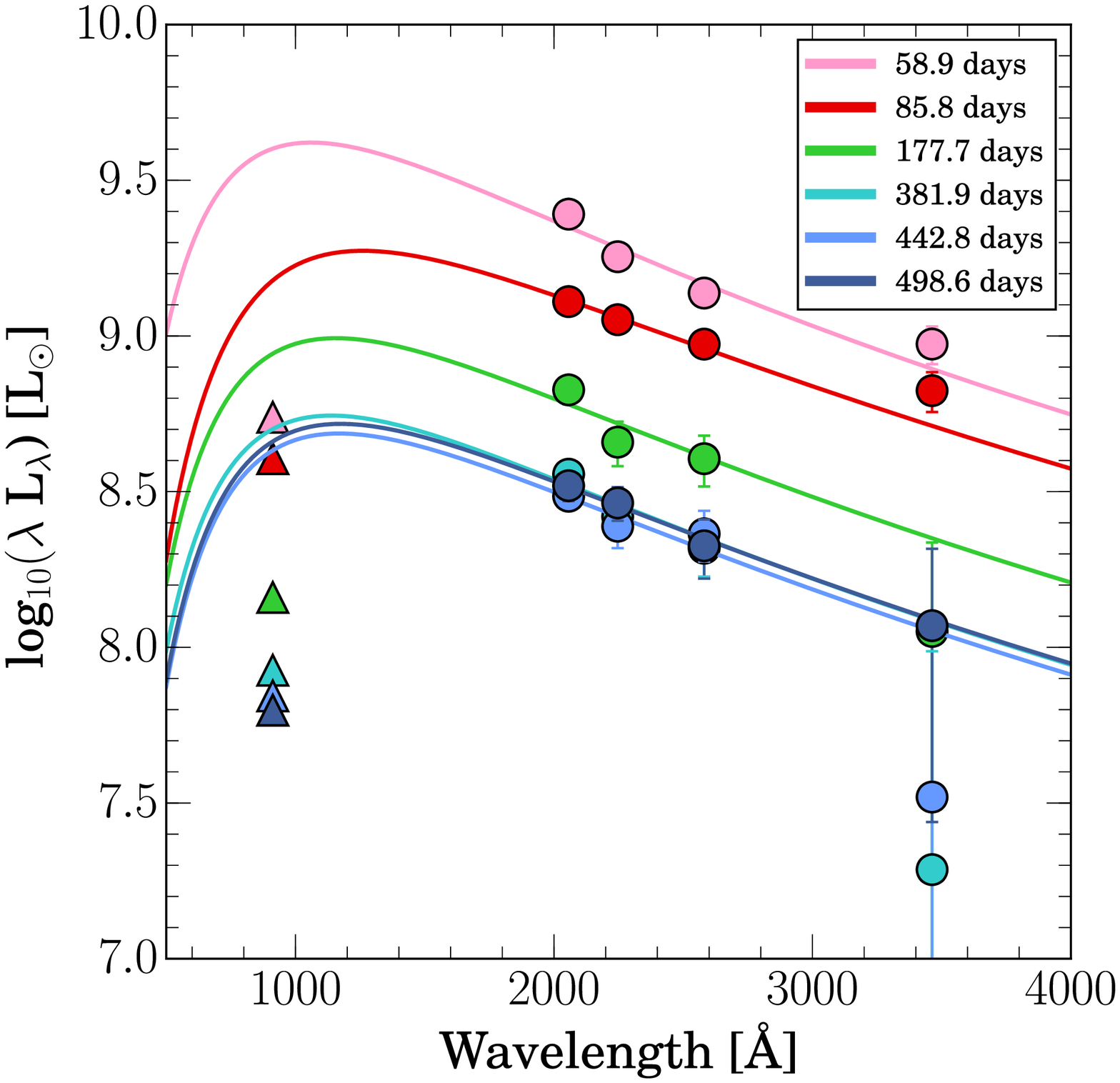}}
\caption{Example blackbody fits to the host-subtracted W2, M2, W1, and U band fluxes. The UVOT epochs were selected to match the dates of our spectroscopic observations; each epoch is assigned a color corresponding to the spectra in Figure~\ref{fig:specEvol}. The solid lines show blackbody fits to the host-subtracted fluxes. The triangles show the lower limits required to produce the observed \halpha\ emission.}
\label{fig:bbSed}
\end{figure}

In Figure~\ref{fig:swiftLC} we show the photometric evolution of \asli\ in the UVOT bands (left axis) as well as in the XRT 0.3--10.0~keV energy range (right axis). In \citet{Holoien14} we presented the first $\sim 175$ days; here we extend our observations to $\sim 600$ days after discovery. The horizontal dashed lines show the UVOT host magnitudes synthesized from the SED fit to the GALEX (NUV), SDSS ($u'$, $g'$, $r'$, $i'$, and  $z'$), and 2MASS ($J$, $H$, $K$) archival data, as well as the upper limit on the X-ray count rate estimated from the archival flux limit of $7.5\times10^{-14}$ ergs s$^{-1}$ cm$^{-2}$ \citep[see][]{Holoien16_14li}. The vertical marks along the time-axis show the dates corresponding to our spectroscopic observations. The extensive archival data (most importantly the GALEX NUV) allows us to constrain the host SED well. We estimate the accuracy of our synthetic host magnitudes with a bootstrapping scheme in which we perturb the input fluxes according to their 1-$\sigma$ errors and fit the resulting SED. We perform 1000 realizations and find that the resulting magnitude estimates are accurate to within $\sim 0.1$ mag.

The most striking aspect of Figure~\ref{fig:swiftLC} is that \asli\ is still bright relative to the host galaxy in the UV and X-ray bands, even after $\sim 600$ days. The excess emission above that of the host decreases toward redder filters, and the photometry in the optical bands becomes consistent with the host $\sim200$ days after discovery.

In order to characterize the excess emission, we correct all fluxes for Galactic extinction assuming $R_V = 3.1$ and $A_V = 0.07$ \citep{Odonnell94,Schlafly11}. We then subtract the host flux and model the SED of the flare as a blackbody using MCMC methods \citep{Foreman-Mackey13}. We fit the W2, M2, W1, and U band fluxes, and exclude the B and V photometry due to the negligible excess flux in those bands. At later epochs, the excess flux in the U band also becomes relatively weak, but the measurements are consistent with the models given the observational uncertainties. Figure~\ref{fig:bbSed} shows example fits to the UV+U band photometry for the dates closest to our spectroscopic observations. The colors denote the observation epoch and correspond to those in Figure~\ref{fig:specEvol}. The circles show the host subtracted fluxes in the W2, M2, W1, and U bands, and the triangles denote lower limits required to produce the observed \halpha\ flux. Not only do we find that the observed UVOT fluxes are reasonably well described by a blackbody, but such a model also produces enough ionizing flux to power the \halpha\ emission at all epochs. 

\begin{figure}
\centering{\includegraphics[scale=1.,width=0.5\textwidth,trim=0.pt 0.pt 0.pt 0.pt,clip]{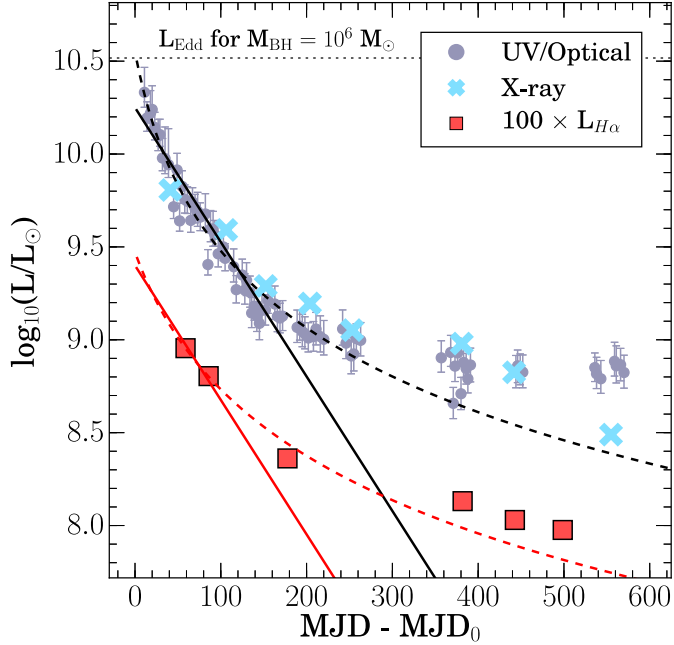}}
\caption{Luminosity evolution of \asli. Gray circles show the bolometric luminosity inferred from our blackbody fits to the host subtracted UVOT fluxes, cyan crosses show the X-ray luminosity in the 0.3--10.0~keV band, and red squares show the measured \halpha\ luminosity scaled upward by a factor of 100 for visibility. The solid black line shows the exponential fit from \citet{Holoien16_14li}, and the solid red line shows the same fit renormalized to the earliest \halpha\ observation. The black and red dashed lines show power-law fits ($\Gamma=-5/3$) to the UV/optical and \halpha\ luminosity, respectively.}
\label{fig:bbLum}
\end{figure}

As we found in \citet{Holoien16_14li}, the UVOT photometry suggests \asli\ evolves at a roughly constant temperature ($\sim3.5\times10^4$ K), while the photospheric radius decreases by a factor of a few over the course of our observations. This inference is driven in part by the fact that we lack the wavelength coverage needed to place a strong constraint on the temperature. We estimate the bolometric luminosity by integrating the fits to the SED of the flare for each of our SWIFT epochs, and derive confidence intervals based on the values containing 68\% of the MCMC distribution for each epoch. We show the evolution of the bolometric luminosity as gray circles in Figure~\ref{fig:bbLum}, along with the evolution of the 0.3--10.0~keV X-ray luminosity (cyan crosses), and a scaled version of the \halpha\ luminosity (red squares).

Figures~\ref{fig:bbLum} shows that there is a marked change in the luminosity evolution of the flare $\sim200$ days after discovery. In \citet{Holoien16_14li} we showed that the initial decline in luminosity is well described by an exponential decay with an $e$-folding time of $\sim60$ days (solid black line). However, after 200 days, the luminosity evolution becomes significantly shallower and only declines by a factor of a few over the next 400 days, causing the exponential to under-predict the luminosity at later times. The dashed black line shows a fit to the UV/optical data assuming $L\propto (t-t_0)^{-5/3}$. We find a best-fit value for $t_0 = -29$ days, which is consistent with our ASAS-SN pre-discovery upper limits. 

We perform a similar exercise for the \halpha\ emission (red squares). The solid red line shows a scaled version the exponential fit from \citet{Holoien16_14li}, and the dashed line shows a power-law fit to the \halpha\ luminosity. While an exponential can provide an adequate fit to the early-time data, it significantly under-predicts the luminosity at later times. Additionally, as we found in \citet{Brown16b}, the \halpha\ luminosity decays on a longer timescale than the bolometric luminosity. The power-law fit to the \halpha\ luminosity prefers a slightly larger value of $t_0 = -56$ days, which is consistent with our ASAS-SN pre-discovery upper limits. While a $\Gamma = -5/3$ power-law fit is an improvement over an exponential for the late-time bolometric and \halpha\ luminosity, \asli\ remains significantly brighter than these power-law models predict. Furthermore, a wide variety of power-law models can provide adequate fits to the data, given our weak pre-discovery constraints.

Based on Eddington arguments and the assumption that the X-ray emitting material be located outside the innermost stable circular orbit, \citet{Miller15} inferred that $\mbh \sim 2 \times 10^6$\msun. Well-established SMBH--host galaxy scaling relations \citep[e.g.][]{Kormendy13,McConnell13} suggest that this may be an underestimate. Based on the host bulge mass \citet{Mendel14}, one would infer $\mbh \sim 10^{6.7}$\msun. However, there is significant intrinsic scatter in these scaling relations \citep{Gultekin09}. Given that the host galaxy may have recently undergone a merger, an episode of significant star formation, and may potentially host a binary SMBH system \citep{Prieto16,RomeroCanizales16}, it is not inconceivable for the SMBH responsible for the TDE to be undermassive relative to the stellar mass of the host galaxy. Interestingly, even for $\mbh = 10^6$\msun, the luminosity of \asli\ remains below Eddington throughout our observing campaign. With these considerations in mind, we adopt a fiducial mass for the SMBH $\mbh = 10^6$\msun.

\begin{figure}
\centering{\includegraphics[scale=1.,width=0.5\textwidth,trim=0.pt 0.pt 0.pt 0.pt,clip]{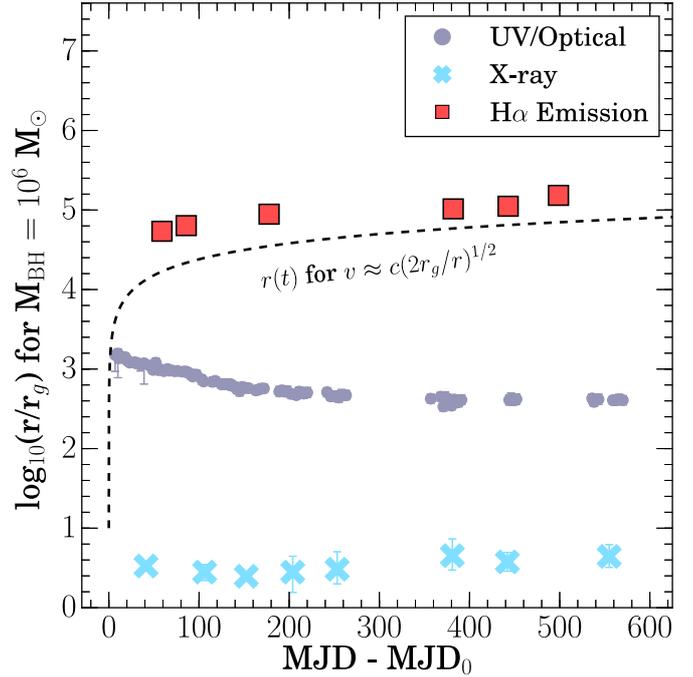}}
\caption{Radius evolution of \asli. Gray circles show the radii inferred from our blackbody fits to the host subtracted UVOT fluxes, the cyan crosses show the characteristic X-ray radius from our blackbody fits to the binned XRT spectra, and red squares show the radii inferred from the \halpha\ line width. All quantities are shown relative to the gravitational radius of the SMBH assuming $\mbh=10^6$\msun. The dashed line shows the evolution of a parabolic orbit with a closest approach equal to the tidal radius.}
\label{fig:bbRad}
\end{figure}

Given that the luminosity decreases with time and the temperature is roughly constant (but only weakly constrained by the UVOT data), the radius must also decrease with time. Figure~\ref{fig:bbRad} shows the radial evolution of various emission components relative to the gravitational radius, where $r_g = G\mbh/c^2$, and we have adopted a fiducial mass $\mbh = 10^6$\msun. Gray circles show the radii inferred from our blackbody fits to the host subtracted UVOT fluxes, the cyan crosses show the characteristic X-ray radius from our blackbody fits to the binned XRT spectra, and red squares show the radii inferred from the \halpha\ line width. The dashed line shows the radial evolution of a parabolic orbit with closest approach equal to the tidal radius. Overall, the photospheric radius (gray circles) decreases, but only by a factor of $\sim2$, and less dramatically than the other ASASSN TDEs \citep{Holoien16_15oi}. The characteristic radius inferred from the UV/optical is on the order of $\sim10$~AU. This corresponds to several tens of tidal radii, but that depends directly on the assumed \mbh. 

We infer a characteristic radius of the line emitting gas (red squares) under the assumption that the width of the \halpha\ feature reflects the characteristic velocity at a given radius, and that $v \approx c(2r_g/r)^{1/2}$. The \halpha\ emission arises from much larger radii than both the UV continuum and X-ray emission (cyan crosses). Additionally, the \halpha\ emitting material appears to be expanding at a rate comparable to that of a parabolic orbit. Given a sufficiently uniform and high cadence dataset, the radial distribution of the emission components could be testable by reverberation mapping to measure the correlations between variability in the X-rays, UVOT passbands, and optical emission lines. Unfortunately, our dataset does not meet the criteria necessary for such an analysis, but reverberation studies in TDEs have recently been attempted \citep[e.g.][]{Jiang16,Kara16}.

\begin{figure}
\centering{\includegraphics[scale=1.,width=0.5\textwidth,trim=0.pt 0.pt 0.pt 0.pt,clip]{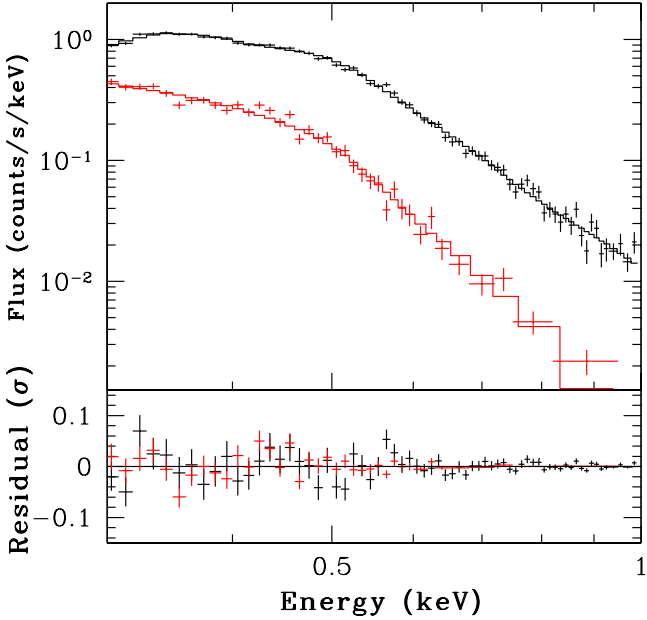}}
\caption{Merged X-ray spectra of \asli\ seen at early times (first 200 days, black crosses) and later times (after 200 days, red crosses). Each spectrum is binned with a minimum of 20 counts per energy bin. The early-time spectrum is best fit with with two Gaussian emission features superimposed on a blackbody (black solid line), while the late-time spectrum is consistent with a purely blackbody model (red solid line). The residuals from the fits are shown in the bottom panel.}
\label{fig:xraySpec}
\end{figure}

\subsection{Evolution of the X-ray Emission}
\label{sec:xray}

Compared to other optically discovered TDEs \citep[e.g.][]{Gezari12,Holoien14,Arcavi14,Holoien16_15oi}, \asli\ is X-ray bright. Over the first $\sim200$ days, the average background subtracted count rate was $\sim0.25$ s$^{-1}$, similar to that derived by \citet{Holoien16_14li}. At later times, the count rate is a factor of 5 lower ($\sim0.05$ s$^{-1}$). However, even though the emission has decreased significantly by late times, Figure~\ref{fig:swiftLC} shows that, even 600 days after discovery, the X-ray flux remains an order of magnitude above the \emph{ROSAT} archival upper limit.

\begin{table}
\begin{minipage}{\textwidth}
\centering
\caption{ASASSN-14li X-ray Properties\hfill}\begin{tabular}{lcccc}
\hline
Mean & kT & Radius & Luminosity \\
MJD & [keV] & [$10^{13}$ cm]  & [$10^{42}$ ergs s$^{-1}]$ \\
\hline
57024.81 & 0.070 $\pm$ 0.001 &  0.49 $\pm$  0.04 & 24.49 $\pm$ 0.29 \\ 
57089.83 & 0.064 $\pm$ 0.002 &  0.41 $\pm$  0.10 & 14.93 $\pm$ 0.39 \\ 
57136.02 & 0.064 $\pm$ 0.001 &  0.37 $\pm$  0.03 &  7.46 $\pm$ 0.12 \\ 
57187.54 & 0.058 $\pm$ 0.003 &  0.41 $\pm$  0.21 &  6.00 $\pm$ 0.17 \\ 
57236.79 & 0.056 $\pm$ 0.005 &  0.45 $\pm$  0.23 &  4.31 $\pm$ 0.15 \\ 
57364.53 & 0.050 $\pm$ 0.004 &  0.67 $\pm$  0.32 &  3.65 $\pm$ 0.11 \\ 
57425.53 & 0.049 $\pm$ 0.003 &  0.56 $\pm$  0.15 &  2.56 $\pm$ 0.17 \\ 
57538.48 & 0.043 $\pm$ 0.003 &  0.65 $\pm$  0.22 &  1.18 $\pm$ 0.13 \\ 
\hline
\end{tabular}
\medskip
\raggedright
\label{tab:xrayProp}
\end{minipage}
\end{table}

We show the X-ray luminosity evolution as cyan crosses in Figure~\ref{fig:bbLum}. Due to the low S/N of the individual epochs, we divide our observations into 8 equally spaced time bins and combine the observations in each bin. While there is short timescale variability, this produces 8 spectra for which we can obtain a robust blackbody fit. Overall, we find that the X-ray luminosity remains comparable to the UV/optical luminosity throughout the 600 day follow-up campaign. The X-ray luminosity at $\sim150$ days presented here is a factor two $\sim2$ lower than that originally found in \citet{Holoien16_14li}. This is mainly due to the fact that we model the data as an absorbed blackbody, which provides a significantly better fit to the observations than the power-law used in \citet{Holoien16_14li}. 

In Figure~\ref{fig:bbRad} we also show the characteristic radius of the X-ray emission (cyan crosses) assuming blackbody emission and a spherical geometry. Viewed as a blackbody, the X-ray emitting surface is on the order of a few gravitational radii, and shows only moderate temporal evolution. We can characterize the overall evolution of the X-ray spectrum by dividing the observations into two sets representing early and late times. The early-time observations corresponds to the first $\sim200$ days of X-rays observations that were analysed by \citet{Holoien16_14li}, while the late time observations correspond to the observations over the remaining $\sim360$ days. The merged early and late time observations have a total exposure times of 131 and 63 ks, respectively.

In Figure~\ref{fig:xraySpec} we show the merged X-ray spectrum obtained for both the early (black) and late (red) time emission. We find that the best fit blackbody model has a temperature of $kT_{\rm early}=0.068\pm0.001$~keV and $kT_{\rm late}=0.056\pm0.002$~keV, and a column density of $N_{H}^{\rm early}=(4.5\pm0.05)\times10^{20}$ cm$^{-2}$ and $N_{H}^{\rm late}=(2.1\pm0.1)\times10^{20}$ cm$^{-2}$ respectively. The total column densities along the line of sight are a factor of 2--3 larger than \Ngal\ \citep{Kalberla05}, and comparable to the values from \citet{Miller15}. However, the reduced chi-squared $\chi_{r}^{2}$ values for these fits are relatively large ($\sim3$ and $\sim1.7$, respectively), suggesting that there are higher order effects not captured by our simple absorbed blackbody model. We also fit the early-time X-ray spectra with two Gaussians superimposed on a blackbody. We find that this significantly improves our fits to the observations. The energies of the Gaussian emission features are centered on 0.376~and~0.406~keV, which are near  strong C and N lines at $\sim0.3675$~and~$0.4307$~keV, respectively. If that is indeed the origin of these features, it would be consistent with the presence of strong, highly ionized C and N emission lines in the UV \citep{Cenko16}. Interestingly, \citet{Miller15} also suggest several highly ionized emission features are present in their high resolution spectra, particularly near the O K-edge.

Irrespective of the origin of the emission features, the X-ray emission from ASASSN-14li is reasonably well described by a blackbody. The temperature of the X-ray emission shows only a moderate decrease with time, from $\sim8\times10^5$~K down to $4.5 \times 10^5$~K. We also find that between the early and late-time observations, the mean column density decreases by a factor of $\sim2$, becoming comparable to \Ngal\ at later times. This could indicate that at early times the material surrounding the TDE was dense, not highly ionized, and obscured the lower energy X-rays, while at late times the material surrounding the event has been nearly completely ionized and thus transparent to the lower energy X-ray photons. We note that the amount of material required to produce this absorbing column is small relative to the overall mass budget, even in the contrived scenario where all of the absorbing material is spherically distributed at the \halpha\ emitting radius: M$_{\rm abs} \sim 0.01 f r^{2}_{17} N_{20} \msun$, where $f$ is the covering fraction, $r_{17}$ is the characteristic radius of the absorbing material in units of $10^{17}$~cm, and $N_{20}$ is the hydrogen column density in units of $10^{20}$~cm$^{-2}$. Furthermore, our modeling suggests there may be a factor of a few variability in the column density between the individual epochs. Unfortunately the spectra for the individual epochs do not reach the S/N required for a precise determination of the column density. We also note similar variability and overall decrease in the absorbing column in the 8 temporally binned spectra. The moderately decreasing temperature and variability in the column density are similar to the findings from \citet{Miller15}, but we have increased the temporal baseline by a factor of $\sim3$.

Finally, we note that the SWIFT XRT detector lacks the energy coverage to discriminate between a bremsstrahlung and blackbody spectrum. While we have assumed the X-ray emission to be well described as a blackbody, there is only marginal evidence that the X-ray spectrum turns over at lower energies. It is plausible that the X-ray emission arises from tidal debris shocking the ambient medium in the vicinity of the SMBH. For ionized gas in the strong shock limit, the temperature is roughly $T \sim 1.4 \times 10^7 v_3^2$ K, where $v_3$ is the shock velocity relative to $10^3$~km~s$^{-1}$ \citep{Draine11}. The characteristic shock velocities needed to produce the observed X-ray temperature are at most a few hundreds of km~s$^{-1}$. The bremsstrahlung emissivity is roughly $\epsilon_{ff} \sim 1.7 \times 10^{-18} T_6^{1/2} n_3^2$ ergs~s$^{-1}$~cm$^{-3}$, where $T_6$ is the temperature relative to 10$^6$~K and $n_3$ is the density relative to 10$^3$~cm$^{-3}$ \citep{Rybicki79}. If the X-rays were to originate only from bremsstrahlung emission, it would require an emitting volume with a characteristic radius of $r_s \sim 10^{20} T_6^{-1/6}n_3^{-2/3}$~cm. Thus the X-ray emission most likely originates from a blackbody-like source powered by accretion onto the SMBH, rather than shocks in the ambient medium. 

\section{Conclusions}
\label{sec:conclusions}

\begin{figure*}
\centering{\includegraphics[scale=1.,width=\textwidth,trim=0.pt 0.pt 0.pt 0.pt,clip]{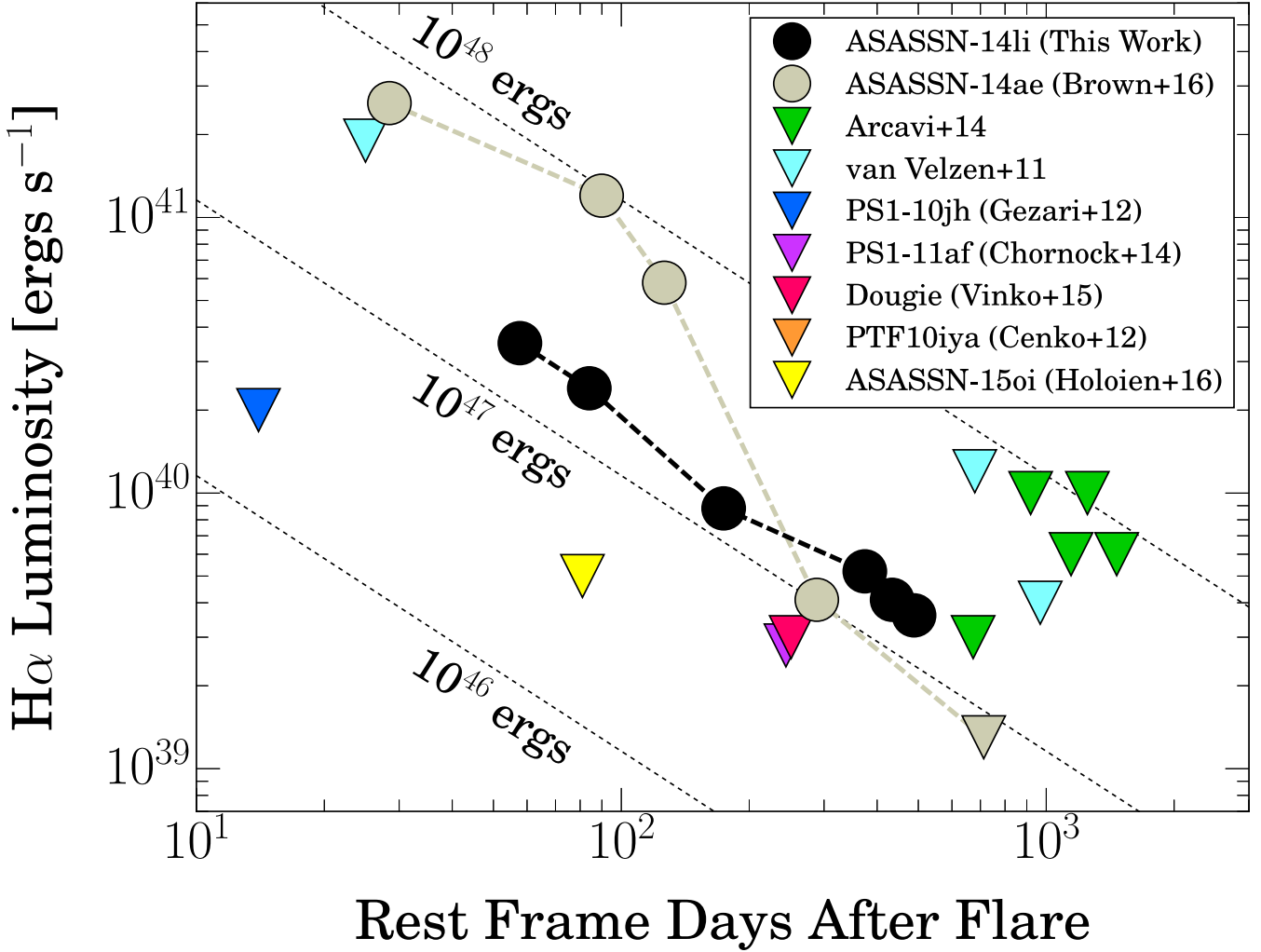}}
\caption{Measurements and limits on late-time \halpha\ emission from optical TDE candidates. The late-time spectra of most of these objects are simply assumed to be host dominated and lack formal upper limit estimates. For these objects, we assume an upper limit of 2\AA\ for the \halpha\ equivalent width, and compute the luminosity based on the approximate continuum and distances to the hosts available for each TDE. The dotted lines show lines of constant emitted energy.}
\label{fig:ensemble}
\end{figure*}

\begin{figure}
\centering{\includegraphics[scale=1.,width=0.5\textwidth,trim=0.pt 0.pt 0.pt 0.pt,clip]{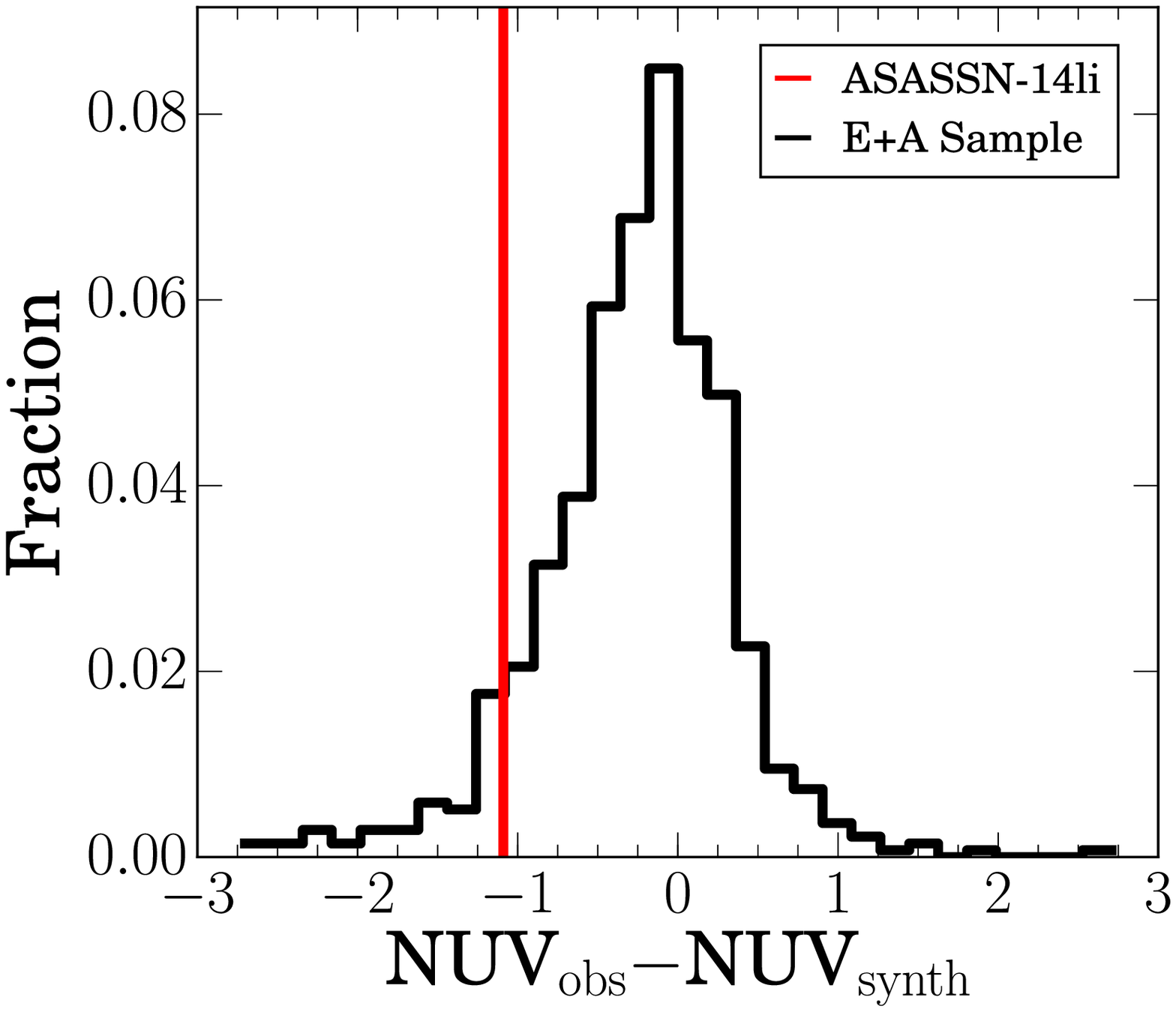}}
\caption{Histogram showing the difference between observed GALEX NUV magnitudes and the expected NUV magnitudes based on models of the optical SED for a sample of \ea\ galaxies from the SDSS. The vertical dashed line shows the difference between the \asli\ late-time NUV and the archival GALEX NUV magnitudes of the host. We estimate the late-time NUV magnitude based on the SWIFT UVW2 observations.}
\label{fig:synthObsComp}
\end{figure}

We have presented late-time optical follow-up spectra taken with LBT/MODS1, extensive UVOT and XRT observations from \SWIFT, and improved ASAS-SN pre-discovery non-detections of the nearby TDE \asli. Our observations span from the epoch of detection to $\sim$~600 days after discovery. In contrast to the late time evolution of ASASSN-14ae \citep{Brown16b}, observations of \asli\ show that TDEs can remain luminous, particularly in the UV and X-ray, for many hundreds of days. We find that the energy radiated in the X-rays is comparable to that of the UV/optical. Integrating over the duration of our campaign, we find that the total radiated energy is $E \approx 7 \times 10^{50}$ ergs. The entire event can be powered by the accretion of a small faction of the overall mass budget ($\Delta M \sim 4 \times 10^{-3}\eta_{0.1}^{-1}$\msun), where $\eta_{0.1}$ is the radiative efficiency relative to 0.1.

While the late-time emission is broadly consistent with the accretion of material onto an SMBH, \asli\ differs from typical AGN. For example, as we have observed in other TDEs (\citealt{Holoien14,Holoien16_14li,Holoien16_15oi} and \citealt{Brown16b}), the optical emission lines narrow as the luminosity decreases, which is the reverse of the behavior observed in AGN \citep{McGill08,Denney09}.

In Figure~\ref{fig:ensemble} we show our late time measurements of the \halpha\ luminosity from \asli. Unlike ASASSN-14ae, the \halpha\ luminosity in \asli\ remains significant for at least 500 days after disruption. We emphasize that the late-time brightness of \asli\ is not simply due to its proximity; it is indeed more luminous at later times than other TDE candidates. Similarly, the evolution of the \halpha\ line width in \asli\ spans a much smaller dynamic range than that of ASASSN-14ae. While the evolution of the \halpha\ line width and luminosity likely encodes information about the evolution of the tidal debris, a larger sample of objects with follow-up spectra is required in order to draw any firm conclusions.

The modest decrease in the X-ray temperature is one of the strongest observational constraints to come out of this work. Most TDE theory predicts that the observed spectrum will become harder at later times \citep[e.g.][]{Lodato11,Strubbe11,Metzger16}, but these claims are typically made in the context of a super-Eddington outflow that obscures the X-rays at early times. For the assumed black hole mass of $10^6\msun$, the X-ray (and optical/UV) luminosity remains below the Eddington limit even at early times, immediately bringing into question the applicability of these theoretical predictions to \asli. Given the variable nature of the X-ray spectra from one epoch to the next as well as the early-time results from \citet{Miller15}, it is likely that the column density and ionization state of the absorbing material along the line of sight are variable, further complicating the evolution of the observed spectrum. This is supported by the fact that X-ray emission comparable to the optical/UV emission is not ubiquitous among optically selected TDE candidates \citep[e.g.][]{Holoien14,Arcavi14,Holoien16_15oi,Brown16b}. Thus, it is not particularly surprising that theoretical predictions fail to match the observations in this instance. We remain agnostic with regard to the adoption of any specific model characterizing the luminosity evolution. While the rate of material returning to pericenter is frequently invoked to explain the luminosity evolution of TDEs, given the complexity of the physical processes involved \citep{Kochanek94,Guillochon13,Guillochon14,Metzger16,Krolik16}, it is likely that the rate of material returning to pericenter has limited bearing on the overall luminosity evolution.

The extended lifetime of \asli\ has important implications for future TDE studies. Our results demonstrate that, even after 500 days, spectra of TDE host galaxies may be contaminated by residual emission from the flare, particularly near strong recombination lines. While this complicates the characterization of TDE host galaxies, it extends the baseline over which residual TDE emission can potentially be discovered in spectroscopic surveys like MaNGA \citep{Bundy15}. Unfortunately, even under the assumption that TDE rates are sharply peaked in E+A galaxies \citep{French16}, the chances of trivially discovering residual TDE emission in a MaNGA-like survey may be hampered by the rarity of these galaxies \citep[see the discussion in][]{Brown16b}.

Alternatively, the \halpha\ emission in TDEs is driven by the ionizing UV continuum, and the strong residual UV emission results in unusually blue UV$-$optical colors. In order to assess the peculiarity of the late time colors, we examine the UV$-$optical colors of a sample of \ea\ galaxies selected from the SDSS. The original selection criteria is described in \citet{Goto04,Goto07}. In short, the selection criteria require that EW(H$\delta$) $> 5.0$\AA, EW([\ion{O}{ii}]) $> -2.5$\AA, EW(H$\alpha$)$ > -3.0$\AA, and S/N$(r) > 10$. We adopt an updated catalog\footnote{\url{http://www.phys.nthu.edu.tw/~tomo/cv/index.html}} based on the SDSS Data Release 7 \citep{Abazajian09}, yielding an initial sample of 837 \ea\ galaxies. We then select a subsample of \ea\ galaxies that also have a GALEX NUV detection within 5\farcs0 of the SDSS position, yielding 683 galaxies with both SDSS optical and GALEX NUV observations. In the absence of late-time GALEX observations of \asli, we estimate the late-time GALEX NUV magnitude based on the late-time SED and find that $NUV \approx W2-0.14$. We find that, relative to other \ea\ galaxies, the late-time NUV$-$optical colors of \asli\ are indeed substantially bluer than the majority of \ea\ galaxies.

Given the peculiarity of the \asli\ late-time colors, we also attempt to identify galaxies with similarly peculiar NUV$-$optical colors. We model each E+A galaxy with the public SED fitting code FAST \citep{Kriek09} based on the SDSS $u,g,r,i,$ and $z$ photometry. We then estimate synthetic GALEX NUV magnitudes from the best fit SED for each galaxy, and compute the difference between the observed GALEX NUV magnitudes and the synthetic NUV magnitudes based on the optical SED. The distribution of the observed and synthetic magnitude difference is shown as the histogram in Figure~\ref{fig:synthObsComp}. The vertical red line shows the late-time NUV excess for \asli, which we know is due to residual TDE emission. We note that the early-time NUV excess for \asli\ (as well as ASASSN-14ae and ASASSN-15oi) is several magnitudes bluer, well outside the range shown here.

We can estimate the upper limit of the TDE rate in \ea\ galaxies by examining the UV excess observed in our sample and making a few basic assumptions. The first assumption we make is that UV excesses larger than that of \asli\ are due to residual emission from a TDE. We also assume that the residual emission is strictly blueward of the optical bandpasses and does not affect the SED modeling of the host galaxy, which is justifiable given the short duration of excess optical TDE continuum emission seen in \asli\ and other TDEs \citep[e.g.][]{Holoien14,Holoien16_15oi}. Additionally, our modeling of the SED is clearly imperfect, as there are a number of galaxies that appear significantly \textit{dimmer} in the UV than the optical colors would suggest. Thus some of the apparent UV excesses are likely due to systematic errors in the modeling. We estimate the magnitude of this contamination by counting the number of galaxies with UV deficits that are greater in magnitude (relative to 0) than the UV excess of \asli, and subtract this number (9) from the population of galaxies with UV excesses greater than that of \asli\ (41). This leaves 32 out 683 \ea\ galaxies showing excess UV emission characteristic of residual TDE emission. If we make the additional assumption that the residual UV emission remains for, on average 2 years, this yields an upper limit on the rate of $\sim 2\times 10^{-2}$ yr$^{-1}$ per galaxy, which is approximately an order of magnitude larger than the rate estimate from \citet{French16}. Alternatively, if no instances of UV excess are associated with residual TDE emission, then the upper limit on the TDE rate in E+A galaxies becomes $\sim7\times10^{-4}$. We note that this is simply an illustrative exercise and is subject to many systematic effects, including the robustness of the SED modeling, potential source confusion in the NUV, and contamination by other sources of UV excess \citep{Oconnell99,Brown04}. A similar exercise with a large sample of early type galaxies yields an even broader distribution in NUV$_{\rm obs}-$NUV$_{\rm synth}$, suggesting that the E+A galaxies with blue excess may indeed be due to modeling systematics. Nonetheless, the substantial UV excess in some of the galaxies may be due to residual emission from a TDE. Identification of residual TDE emission in spectroscopic surveys like MaNGA is reliant upon archival spectra of the host, whereas the approach illustrated here has yielded a relatively small sample of galaxies well suited for targeted follow-up observations without the need for archival spectroscopy. Spectroscopic signatures that are sensitive to TDE emission on a longer temporal baseline \citep[e.g. coronal line emission][]{Komossa08,Wang11,Yang13} could prove useful in determining if the UV excess in these galaxies is due to residual TDE emission and, ultimately, improving our understanding of TDE host galaxies.

\section*{Acknowledgements}

JSB, KZS, and CSK are supported by NSF grants AST-1515876 and AST-1515927.

TW-SH is supported by the DOE Computational Science Graduate Fellowship, grant number DE-FG02-97ER25308.

BJS is supported by NASA through Hubble Fellowship grant HF-51348.001 awarded by the Space Telescope Science Institute, which is operated by the Association of Universities for Research in Astronomy, Inc., for NASA, under contract NAS 5-26555. 

Support for JLP is provided in part by FONDECYT through the grant 1151445 and by the Ministry of Economy, Development, and Tourism's Millennium Science Initiative through grant IC120009, awarded to The Millennium Institute of Astrophysics, MAS.

This paper used data obtained with the MODS spectrographs built with funding from NSF grant AST-9987045 and the NSF Telescope System Instrumentation Program (TSIP), with additional funds from the Ohio Board of Regents and the Ohio State University Office of Research.

Based on data acquired using the Large Binocular Telescope (LBT). The LBT is an international collaboration among institutions in the United States, Italy, and Germany. LBT Corporation partners are: The University of Arizona on behalf of the Arizona university system; Istituto Nazionale di Astrofisica, Italy; LBT Beteiligungsgesellschaft, Germany, representing the Max-Planck Society, the Astrophysical Institute Potsdam, and Heidelberg University; The Ohio State University, and The Research Corporation, on behalf of The University of Notre Dame, University of Minnesota and University of Virginia.

This research has made use of the XRT Data Analysis Software (XRTDAS) developed under the responsibility of the ASI Science Data Center (ASDC), Italy. At Penn State the NASA \SWIFT\ program is supported through contract NAS5-00136.

\bibliography{late14li}
\bsp	

\appendix
\section{Follow-up Photometry}
All UVOT and XRT follow-up photometry is presented in Table~\ref{tab:phot} below. UVOT photometry is presented in the Vega system, and XRT photometry is presented in counts~s$^{-1}$. The data have not been corrected for Galactic absorption.

\begin{table*}
\begin{minipage}{\textwidth}
\caption{\SWIFT\ Observations.\hfill}\begin{tabular}{cccccccc}
\hline
MJD & XRT counts s$^{-1}$ & W2 & M2 & W1 & U & B & V \\
\hline
56991 & 0.316 $\pm$ 0.011  & 14.15 $\pm$ 0.04 & 14.39 $\pm$ 0.03 & 14.66 $\pm$ 0.05 & 15.14 $\pm$ 0.04 & 15.97 $\pm$ 0.04 & 15.55 $\pm$ 0.06 \\
56994 & 0.408 $\pm$ 0.013  & 14.59 $\pm$ 0.04 & 14.84 $\pm$ 0.03 & 14.91 $\pm$ 0.05 & 15.26 $\pm$ 0.04 & 16.03 $\pm$ 0.04 & 15.65 $\pm$ 0.06 \\
56995 & 0.318 $\pm$ 0.011  & 14.20 $\pm$ 0.03 & 14.47 $\pm$ 0.03 & 14.75 $\pm$ 0.04 & 15.22 $\pm$ 0.04 & 15.93 $\pm$ 0.04 & 15.56 $\pm$ 0.05 \\
56998 & 0.335 $\pm$ 0.011  & 14.25 $\pm$ 0.05 & 14.60 $\pm$ 0.03 & 14.81 $\pm$ 0.05 & 15.25 $\pm$ 0.05 & 15.94 $\pm$ 0.04 & 15.60 $\pm$ 0.06 \\
57002 & 0.367 $\pm$ 0.012  & 14.34 $\pm$ 0.03 & 14.60 $\pm$ 0.03 & 14.81 $\pm$ 0.05 & 15.27 $\pm$ 0.04 & 16.01 $\pm$ 0.04 & 15.46 $\pm$ 0.05 \\
57004 & 0.356 $\pm$ 0.018  & 14.37 $\pm$ 0.04 & 14.65 $\pm$ 0.04 & 14.87 $\pm$ 0.05 & 15.39 $\pm$ 0.05 & 15.98 $\pm$ 0.04 & 15.68 $\pm$ 0.10 \\
57007 & 0.452 $\pm$ 0.013  & 14.58 $\pm$ 0.04 & 14.83 $\pm$ 0.03 & 15.08 $\pm$ 0.05 & 15.47 $\pm$ 0.05 & 16.05 $\pm$ 0.04 & 15.63 $\pm$ 0.06 \\
57011 & 0.508 $\pm$ 0.014  & 14.66 $\pm$ 0.04 & 14.86 $\pm$ 0.03 & 15.17 $\pm$ 0.05 & 15.48 $\pm$ 0.05 & 16.04 $\pm$ 0.04 & 15.62 $\pm$ 0.06 \\
57013 & 0.449 $\pm$ 0.013  & 14.62 $\pm$ 0.04 & 14.87 $\pm$ 0.03 & 15.13 $\pm$ 0.05 & 15.43 $\pm$ 0.04 & 16.06 $\pm$ 0.04 & 15.69 $\pm$ 0.06 \\
57016 & 0.310 $\pm$ 0.011  & 14.81 $\pm$ 0.04 & 15.06 $\pm$ 0.03 & 15.27 $\pm$ 0.05 & 15.53 $\pm$ 0.05 & 16.17 $\pm$ 0.04 & 15.67 $\pm$ 0.06 \\
57020 & 0.429 $\pm$ 0.012  & 14.90 $\pm$ 0.04 & 15.09 $\pm$ 0.03 & 15.33 $\pm$ 0.05 & 15.59 $\pm$ 0.05 & 16.16 $\pm$ 0.05 & 15.71 $\pm$ 0.07 \\
57023 & 0.300 $\pm$ 0.011  & 14.87 $\pm$ 0.04 & 15.10 $\pm$ 0.03 & 15.32 $\pm$ 0.05 & 15.53 $\pm$ 0.05 & 16.09 $\pm$ 0.04 & 15.67 $\pm$ 0.05 \\
57029 & 0.426 $\pm$ 0.011  & 15.27 $\pm$ 0.05 & 15.50 $\pm$ 0.05 & 15.54 $\pm$ 0.06 & 15.74 $\pm$ 0.06 & 16.18 $\pm$ 0.06 & 15.67 $\pm$ 0.08 \\
57033 & 0.470 $\pm$ 0.018  & 15.06 $\pm$ 0.04 & 15.29 $\pm$ 0.04 & 15.62 $\pm$ 0.06 & 15.79 $\pm$ 0.07 & 16.16 $\pm$ 0.06 & 15.76 $\pm$ 0.08 \\
57036 & 0.420 $\pm$ 0.015  & 15.45 $\pm$ 0.05 & 15.40 $\pm$ 0.04 & 15.56 $\pm$ 0.05 & 15.77 $\pm$ 0.06 & 16.31 $\pm$ 0.06 & 15.87 $\pm$ 0.08 \\
57039 & 0.380 $\pm$ 0.012  & 15.23 $\pm$ 0.04 & 15.41 $\pm$ 0.04 & 15.66 $\pm$ 0.06 & 15.86 $\pm$ 0.06 & 16.29 $\pm$ 0.05 & 15.85 $\pm$ 0.07 \\
57042 & 0.387 $\pm$ 0.015  & 15.29 $\pm$ 0.05 & 15.54 $\pm$ 0.05 & 15.67 $\pm$ 0.07 & 15.79 $\pm$ 0.07 & 16.19 $\pm$ 0.06 & 15.91 $\pm$ 0.09 \\
57046 & 0.386 $\pm$ 0.014  & 15.39 $\pm$ 0.04 & 15.54 $\pm$ 0.04 & 15.74 $\pm$ 0.06 & 15.87 $\pm$ 0.06 & 16.22 $\pm$ 0.05 & 15.77 $\pm$ 0.07 \\
57049 & 0.366 $\pm$ 0.013  & 15.48 $\pm$ 0.05 & 15.64 $\pm$ 0.04 & 15.74 $\pm$ 0.06 & 15.85 $\pm$ 0.06 & 16.30 $\pm$ 0.06 & 15.76 $\pm$ 0.08 \\
57051 & 0.322 $\pm$ 0.012  & 15.32 $\pm$ 0.05 & 15.58 $\pm$ 0.04 & 15.70 $\pm$ 0.06 & 15.86 $\pm$ 0.06 & 16.36 $\pm$ 0.06 & 15.73 $\pm$ 0.07 \\
57054 & 0.307 $\pm$ 0.011  & 15.43 $\pm$ 0.05 & 15.59 $\pm$ 0.05 & 15.70 $\pm$ 0.07 & 15.95 $\pm$ 0.08 & 16.26 $\pm$ 0.06 & 15.73 $\pm$ 0.08 \\
57058 & 0.316 $\pm$ 0.011  & 15.45 $\pm$ 0.04 & 15.65 $\pm$ 0.04 & 15.69 $\pm$ 0.05 & 15.94 $\pm$ 0.05 & 16.23 $\pm$ 0.04 & 15.75 $\pm$ 0.06 \\
57060 & 0.311 $\pm$ 0.012  & 15.46 $\pm$ 0.05 & 15.70 $\pm$ 0.05 & 15.80 $\pm$ 0.07 & 15.87 $\pm$ 0.07 & 16.32 $\pm$ 0.06 & 15.81 $\pm$ 0.09 \\
57066 & 0.217 $\pm$ 0.011  & 15.43 $\pm$ 0.08 & 15.61 $\pm$ 0.07 & 15.84 $\pm$ 0.06 & 15.95 $\pm$ 0.07 & 16.27 $\pm$ 0.05 & 15.72 $\pm$ 0.07 \\
57069 & 0.247 $\pm$ 0.013  & 15.94 $\pm$ 0.05 & 15.99 $\pm$ 0.04 & 16.00 $\pm$ 0.06 & 15.99 $\pm$ 0.06 & 16.27 $\pm$ 0.04 & 15.79 $\pm$ 0.06 \\
57072 & 0.274 $\pm$ 0.011  & 15.62 $\pm$ 0.04 & 15.79 $\pm$ 0.04 & 15.85 $\pm$ 0.05 & 15.94 $\pm$ 0.05 & 16.34 $\pm$ 0.05 & 15.73 $\pm$ 0.06 \\
57075 & 0.250 $\pm$ 0.010  & 15.65 $\pm$ 0.05 & 15.86 $\pm$ 0.05 & 15.98 $\pm$ 0.06 & 15.99 $\pm$ 0.06 & 16.27 $\pm$ 0.05 & 15.78 $\pm$ 0.08 \\
57078 & 0.137 $\pm$ 0.008  & 15.68 $\pm$ 0.05 & 15.93 $\pm$ 0.04 & 16.08 $\pm$ 0.06 & 16.02 $\pm$ 0.07 & 16.33 $\pm$ 0.05 & 15.83 $\pm$ 0.08 \\
57081 & 0.224 $\pm$ 0.009  & 15.82 $\pm$ 0.07 & 16.04 $\pm$ 0.07 & 16.03 $\pm$ 0.09 & 16.04 $\pm$ 0.10 & 16.35 $\pm$ 0.08 & 15.69 $\pm$ 0.10 \\
57087 & 0.280 $\pm$ 0.012  & 15.91 $\pm$ 0.04 & 16.05 $\pm$ 0.04 & 16.17 $\pm$ 0.06 & 16.17 $\pm$ 0.07 & 16.34 $\pm$ 0.05 & 15.83 $\pm$ 0.07 \\
57089 & 0.253 $\pm$ 0.010  & 16.00 $\pm$ 0.04 & 16.21 $\pm$ 0.04 & 16.26 $\pm$ 0.06 & 16.23 $\pm$ 0.06 & 16.35 $\pm$ 0.04 & 15.74 $\pm$ 0.06 \\
57099 & 0.279 $\pm$ 0.015  & 16.12 $\pm$ 0.05 & 16.22 $\pm$ 0.06 & 16.37 $\pm$ 0.06 & 16.18 $\pm$ 0.06 & \ldots & \ldots \\
57102 & 0.257 $\pm$ 0.012  & 16.23 $\pm$ 0.07 & 16.51 $\pm$ 0.09 & 16.35 $\pm$ 0.09 & 16.17 $\pm$ 0.07 & \ldots & \ldots \\
57105 & 0.123 $\pm$ 0.013  & \ldots & 16.36 $\pm$ 0.06 & 16.61 $\pm$ 0.14 & \ldots & \ldots & \ldots \\
57109 & 0.244 $\pm$ 0.011  & 16.19 $\pm$ 0.06 & 16.40 $\pm$ 0.08 & 16.43 $\pm$ 0.08 & 16.26 $\pm$ 0.07 & \ldots & \ldots \\
57112 & 0.213 $\pm$ 0.010  & 16.32 $\pm$ 0.05 & 16.50 $\pm$ 0.06 & 16.42 $\pm$ 0.06 & 16.27 $\pm$ 0.05 & \ldots & \ldots \\
57114 & 0.221 $\pm$ 0.013  & 16.23 $\pm$ 0.06 & 16.48 $\pm$ 0.08 & 16.36 $\pm$ 0.09 & 16.33 $\pm$ 0.08 & \ldots & \ldots \\
57118 & 0.183 $\pm$ 0.010  & 16.35 $\pm$ 0.06 & 16.47 $\pm$ 0.06 & 16.39 $\pm$ 0.06 & 16.35 $\pm$ 0.05 & \ldots & \ldots \\
57120 & 0.182 $\pm$ 0.010  & 16.57 $\pm$ 0.06 & 16.61 $\pm$ 0.06 & 16.54 $\pm$ 0.07 & 16.34 $\pm$ 0.06 & \ldots & \ldots \\
57124 & 0.169 $\pm$ 0.009  & 16.50 $\pm$ 0.05 & 16.58 $\pm$ 0.06 & 16.56 $\pm$ 0.07 & 16.37 $\pm$ 0.05 & \ldots & \ldots \\
57126 & 0.166 $\pm$ 0.009  & 16.64 $\pm$ 0.06 & 16.69 $\pm$ 0.08 & 16.66 $\pm$ 0.09 & 16.43 $\pm$ 0.07 & \ldots & \ldots \\
57129 & 0.176 $\pm$ 0.011  & 16.71 $\pm$ 0.08 & 16.72 $\pm$ 0.08 & 16.80 $\pm$ 0.09 & 16.40 $\pm$ 0.07 & \ldots & \ldots \\
57132 & 0.129 $\pm$ 0.008  & 16.53 $\pm$ 0.07 & 16.65 $\pm$ 0.08 & 16.57 $\pm$ 0.09 & 16.52 $\pm$ 0.09 & 16.39 $\pm$ 0.06 & 15.69 $\pm$ 0.08 \\
57136 & 0.139 $\pm$ 0.010  & 16.53 $\pm$ 0.07 & 16.70 $\pm$ 0.10 & 16.60 $\pm$ 0.09 & 16.47 $\pm$ 0.08 & \ldots & \ldots \\
57139 & 0.144 $\pm$ 0.009  & 16.46 $\pm$ 0.05 & 16.58 $\pm$ 0.06 & 16.65 $\pm$ 0.07 & 16.45 $\pm$ 0.05 & \ldots & \ldots \\
57148 & 0.167 $\pm$ 0.009  & 16.54 $\pm$ 0.05 & 16.70 $\pm$ 0.07 & 16.67 $\pm$ 0.07 & 16.47 $\pm$ 0.06 & \ldots & \ldots \\
57150 & 0.182 $\pm$ 0.009  & 16.60 $\pm$ 0.05 & 16.76 $\pm$ 0.06 & 16.68 $\pm$ 0.06 & 16.33 $\pm$ 0.05 & \ldots & \ldots \\
57153 & 0.178 $\pm$ 0.009  & 16.66 $\pm$ 0.05 & 16.66 $\pm$ 0.07 & 16.65 $\pm$ 0.07 & 16.44 $\pm$ 0.06 & \ldots & \ldots \\
57156 & 0.154 $\pm$ 0.010  & 16.55 $\pm$ 0.09 & 16.77 $\pm$ 0.12 & 16.65 $\pm$ 0.12 & 16.54 $\pm$ 0.10 & \ldots & \ldots \\
57173 & 0.157 $\pm$ 0.010  & 16.71 $\pm$ 0.06 & 16.89 $\pm$ 0.09 & 16.71 $\pm$ 0.08 & 16.44 $\pm$ 0.07 & \ldots & \ldots \\
57176 & 0.147 $\pm$ 0.009  & 16.71 $\pm$ 0.05 & 16.82 $\pm$ 0.07 & 16.72 $\pm$ 0.07 & 16.32 $\pm$ 0.06 & \ldots & \ldots \\
\hline
\end{tabular}
\end{minipage}
\end{table*}
\begin{table*}
\begin{minipage}{\textwidth}
\begin{tabular}{cccccccc}
\hline
MJD & XRT counts s$^{-1}$ & W2 & M2 & W1 & U & B & V \\
\hline
57179 & 0.149 $\pm$ 0.016  & 16.73 $\pm$ 0.11 & 16.85 $\pm$ 0.09 & 16.72 $\pm$ 0.14 & 16.33 $\pm$ 0.09 & \ldots & \ldots \\
57182 & 0.147 $\pm$ 0.012  & 16.75 $\pm$ 0.12 & 16.92 $\pm$ 0.17 & 16.71 $\pm$ 0.15 & 16.36 $\pm$ 0.12 & \ldots & \ldots \\
57186 & 0.138 $\pm$ 0.010  & 16.82 $\pm$ 0.07 & 16.96 $\pm$ 0.09 & 16.82 $\pm$ 0.09 & 16.61 $\pm$ 0.07 & \ldots & \ldots \\
57189 & 0.163 $\pm$ 0.034  & \ldots & 17.12 $\pm$ 0.11 & \ldots & \ldots & \ldots & \ldots \\
57192 & 0.141 $\pm$ 0.009  & 16.85 $\pm$ 0.06 & 16.92 $\pm$ 0.07 & 16.91 $\pm$ 0.07 & 16.39 $\pm$ 0.05 & \ldots & \ldots \\
57195 & 0.144 $\pm$ 0.008  & 16.73 $\pm$ 0.09 & 16.86 $\pm$ 0.11 & 16.73 $\pm$ 0.11 & 16.55 $\pm$ 0.09 & \ldots & \ldots \\
57200 & 0.125 $\pm$ 0.010  & 16.83 $\pm$ 0.06 & 16.97 $\pm$ 0.07 & 16.71 $\pm$ 0.07 & 16.50 $\pm$ 0.05 & \ldots & \ldots \\
57204 & 0.112 $\pm$ 0.008  & 16.78 $\pm$ 0.08 & 17.04 $\pm$ 0.11 & 16.74 $\pm$ 0.10 & 16.53 $\pm$ 0.08 & \ldots & \ldots \\
57226 & 0.083 $\pm$ 0.007  & 16.74 $\pm$ 0.07 & 16.89 $\pm$ 0.10 & 16.78 $\pm$ 0.10 & 16.58 $\pm$ 0.09 & \ldots & \ldots \\
57230 & 0.099 $\pm$ 0.011  & 16.88 $\pm$ 0.06 & 17.03 $\pm$ 0.08 & 16.95 $\pm$ 0.09 & 16.54 $\pm$ 0.07 & \ldots & \ldots \\
57236 & 0.094 $\pm$ 0.007  & 16.89 $\pm$ 0.13 & 17.19 $\pm$ 0.14 & 16.85 $\pm$ 0.12 & 16.69 $\pm$ 0.11 & \ldots & \ldots \\
57239 & 0.107 $\pm$ 0.007  & 16.90 $\pm$ 0.11 & 17.13 $\pm$ 0.14 & 17.03 $\pm$ 0.14 & 16.73 $\pm$ 0.13 & \ldots & \ldots \\
57242 & 0.100 $\pm$ 0.007  & 16.87 $\pm$ 0.06 & 16.92 $\pm$ 0.07 & 16.91 $\pm$ 0.08 & 16.51 $\pm$ 0.06 & \ldots & \ldots \\
57247 & 0.074 $\pm$ 0.006  & 16.81 $\pm$ 0.06 & 17.05 $\pm$ 0.08 & 16.92 $\pm$ 0.08 & 16.59 $\pm$ 0.07 & \ldots & \ldots \\
57341 & 0.061 $\pm$ 0.005  & 16.99 $\pm$ 0.06 & 17.13 $\pm$ 0.06 & 17.05 $\pm$ 0.08 & 16.53 $\pm$ 0.07 & 16.44 $\pm$ 0.05 & 15.82 $\pm$ 0.07 \\
57352 & 0.055 $\pm$ 0.005  & 17.01 $\pm$ 0.06 & 17.00 $\pm$ 0.07 & 16.89 $\pm$ 0.07 & 16.51 $\pm$ 0.06 & \ldots & \ldots \\
57355 & 0.049 $\pm$ 0.005  & 17.47 $\pm$ 0.06 & 17.34 $\pm$ 0.07 & 17.18 $\pm$ 0.08 & 16.72 $\pm$ 0.06 & \ldots & \ldots \\
57357 & 0.062 $\pm$ 0.005  & 17.13 $\pm$ 0.06 & 17.12 $\pm$ 0.08 & 16.82 $\pm$ 0.07 & 16.64 $\pm$ 0.06 & \ldots & \ldots \\
57360 & 0.075 $\pm$ 0.007  & 16.97 $\pm$ 0.06 & 17.11 $\pm$ 0.09 & 17.11 $\pm$ 0.10 & 16.69 $\pm$ 0.07 & \ldots & \ldots \\
57364 & 0.087 $\pm$ 0.006  & 17.24 $\pm$ 0.06 & 17.61 $\pm$ 0.09 & 17.08 $\pm$ 0.08 & 16.57 $\pm$ 0.06 & \ldots & \ldots \\
57367 & 0.076 $\pm$ 0.006  & 17.06 $\pm$ 0.06 & 17.17 $\pm$ 0.07 & 17.03 $\pm$ 0.07 & 16.67 $\pm$ 0.06 & \ldots & \ldots \\
57370 & 0.062 $\pm$ 0.005  & 17.14 $\pm$ 0.06 & 17.21 $\pm$ 0.08 & 17.19 $\pm$ 0.08 & 16.60 $\pm$ 0.06 & \ldots & \ldots \\
57372 & 0.050 $\pm$ 0.005  & 17.27 $\pm$ 0.07 & 17.18 $\pm$ 0.08 & 17.06 $\pm$ 0.08 & 16.57 $\pm$ 0.06 & \ldots & \ldots \\
57376 & 0.065 $\pm$ 0.006  & 17.08 $\pm$ 0.07 & 17.17 $\pm$ 0.08 & 17.10 $\pm$ 0.09 & 16.59 $\pm$ 0.06 & \ldots & \ldots \\
57378 & 0.074 $\pm$ 0.007  & 17.08 $\pm$ 0.07 & \ldots & \ldots & 16.58 $\pm$ 0.07 & \ldots & \ldots \\
57383 & 0.031 $\pm$ 0.008  & \ldots & 17.03 $\pm$ 0.08 & 17.02 $\pm$ 0.09 & 16.55 $\pm$ 0.20 & \ldots & \ldots \\
57411 & 0.054 $\pm$ 0.006  & 17.21 $\pm$ 0.09 & 17.22 $\pm$ 0.13 & 17.26 $\pm$ 0.13 & 16.60 $\pm$ 0.09 & \ldots & \ldots \\
57417 & 0.064 $\pm$ 0.006  & 17.28 $\pm$ 0.10 & 17.13 $\pm$ 0.12 & 17.21 $\pm$ 0.13 & 16.65 $\pm$ 0.09 & \ldots & \ldots \\
57423 & 0.047 $\pm$ 0.009  & 17.34 $\pm$ 0.11 & 17.13 $\pm$ 0.15 & 17.12 $\pm$ 0.14 & 16.57 $\pm$ 0.10 & \ldots & \ldots \\
57426 & 0.029 $\pm$ 0.009  & 17.51 $\pm$ 0.27 & 17.08 $\pm$ 0.10 & 17.19 $\pm$ 0.11 & 16.79 $\pm$ 0.09 & \ldots & \ldots \\
57428 & 0.046 $\pm$ 0.005  & 17.18 $\pm$ 0.06 & 17.22 $\pm$ 0.08 & 16.98 $\pm$ 0.08 & 16.65 $\pm$ 0.06 & \ldots & \ldots \\
57430 & 0.026 $\pm$ 0.007  & 17.12 $\pm$ 0.08 & 17.19 $\pm$ 0.10 & 16.86 $\pm$ 0.09 & 16.78 $\pm$ 0.08 & \ldots & \ldots \\
57433 & 0.040 $\pm$ 0.005  & 17.13 $\pm$ 0.07 & 17.20 $\pm$ 0.09 & 17.12 $\pm$ 0.09 & 16.62 $\pm$ 0.06 & \ldots & \ldots \\
57436 & 0.046 $\pm$ 0.005  & 17.21 $\pm$ 0.06 & 17.06 $\pm$ 0.07 & 17.07 $\pm$ 0.08 & 16.53 $\pm$ 0.05 & \ldots & \ldots \\
57520 & 0.025 $\pm$ 0.004  & 17.13 $\pm$ 0.07 & 17.10 $\pm$ 0.07 & 17.03 $\pm$ 0.09 & 16.53 $\pm$ 0.08 & 16.42 $\pm$ 0.06 & 15.95 $\pm$ 0.08 \\
57523 & 0.019 $\pm$ 0.003  & 17.15 $\pm$ 0.07 & 17.21 $\pm$ 0.06 & 17.19 $\pm$ 0.09 & 16.64 $\pm$ 0.07 & 16.46 $\pm$ 0.05 & 15.88 $\pm$ 0.07 \\
57527 & 0.021 $\pm$ 0.003  & 17.20 $\pm$ 0.07 & 17.25 $\pm$ 0.07 & 16.89 $\pm$ 0.08 & 16.57 $\pm$ 0.08 & 16.54 $\pm$ 0.06 & 15.81 $\pm$ 0.07 \\
57543 & 0.022 $\pm$ 0.004  & 17.06 $\pm$ 0.07 & 17.10 $\pm$ 0.10 & 17.19 $\pm$ 0.10 & 16.55 $\pm$ 0.07 & \ldots & \ldots \\
57545 & 0.018 $\pm$ 0.007  & 17.05 $\pm$ 0.09 & 17.22 $\pm$ 0.13 & 17.09 $\pm$ 0.12 & 16.65 $\pm$ 0.09 & \ldots & \ldots \\
57546 & 0.022 $\pm$ 0.004  & 17.01 $\pm$ 0.08 & 17.18 $\pm$ 0.11 & 17.20 $\pm$ 0.11 & 16.55 $\pm$ 0.08 & \ldots & \ldots \\
57550 & 0.034 $\pm$ 0.005  & 17.02 $\pm$ 0.07 & 17.30 $\pm$ 0.10 & 17.00 $\pm$ 0.09 & 16.42 $\pm$ 0.06 & \ldots & \ldots \\
57554 & 0.012 $\pm$ 0.003  & 17.16 $\pm$ 0.06 & 17.23 $\pm$ 0.08 & 17.02 $\pm$ 0.07 & 16.62 $\pm$ 0.06 & \ldots & \ldots \\
\hline
\end{tabular}
\medskip

\raggedright
\noindent Magnitudes and uncertainties are presented in the Vega system. X-ray count rates anduncertainties are given in units of \\counts per second in the energy range $0.3-10$ keV. Uncertainties are given next to each measurement. Data are not \\ corrected for Galactic absorption.
\label{tab:phot}
\end{minipage}
\end{table*}

\label{lastpage}

\end{document}